\documentclass[a4paper,11pt]{article}
\pdfoutput=1
\usepackage{jcappub}
\usepackage{rotating}
\usepackage{graphicx}
\usepackage{epsfig}
\usepackage{amsmath}
\usepackage{amssymb}
\usepackage{subfigure}
\usepackage{time}
\usepackage{relsize}
\usepackage{xcolor}
\usepackage{amsmath,amssymb}
\usepackage{tensor}
\usepackage[utf8]{inputenc}
\usepackage{cleveref}
\definecolor{blue-violet}{rgb}{0.54, 0.17, 0.89}
\hypersetup{colorlinks=true, citecolor=blue, linkcolor=magenta,
urlcolor = blue}

\def\be{\begin{equation}}
\def\ee{\end{equation}}
\newcommand{\bea}{\begin{eqnarray}}
	\newcommand{\eea}{\end{eqnarray}}

\title{\boldmath No-Hair Theorem in the Wake of Event Horizon Telescope}

\affiliation[a]{School of Astronomy, Institute for Research in Fundamental Sciences (IPM)	P. O. Box 19395-5531, Tehran, Iran}
\affiliation[b]{Dipartimento di Fisica ``E.R Caianiello", Universit$\grave{a}$ degli Studi di Salerno, Via Giovanni Paolo II, 132-84084 Fisciano (SA), Italy
}
\affiliation[c]{Istituto Nazionale di Fisica Nucleare - Gruppo Collegato di Salerno - Sezione di Napoli, Via Giovanni Paolo II, 132 - 84084 Fisciano (SA), Italy}
\affiliation[d]{Institute of Theoretical Astrophysics, University of Oslo, P.O. Box 1029 Blindern, N-0315 Oslo, Norway}

\author[a]{Mohsen Khodadi,}\emailAdd{m.khodadi@ipm.ir}
\author[b,c]{Gaetano Lambiase,}\emailAdd{lambiase@sa.infn.it}
\author[d]{David F. Mota,}\emailAdd{d.f.mota@astro.uio.no}

\abstract
{
	Thanks to the release of the extraordinary  EHT image of shadow attributed to the M87* supermassive black hole (SMBH), we have a novel window to assess the validity of fundamental physics in the strong-field regime. Motivated by this, we consider Johannsen \& Psaltis metric parameterized by mass, spin, and an additional  dimensionless hair parameter $\epsilon$. This parametric framework in the high rotation regimes provides a well-behaved bed to the strong-gravity test of the no-hair theorem (NHT) using the EHT data. Incorporating the $\epsilon$ into the standard Kerr spacetime enrich it in the sense that, depending on setting the positive and negative values for that, we deal with alternative compact objects: deformed Kerr naked singularity and Kerr BH solutions, respectively. Shadows associated with these two possible solutions indicate that the deformation parameter $\epsilon$ affects the geometry shape of standard shadow such that it becomes more oblate and prolate with $\epsilon<0$ and $\epsilon>0$, respectively. By scanning the window associated with three shadow observables oblateness, deviation from circularity, and shadow diameter, we perform a numerical analysis within the range $a_*=0.9\mp0.1$ of the dimensionless rotation parameter, to find the constraints on the hair parameter $\epsilon$ in both possible solutions. For both possible signs of $\epsilon$,  we extract a variety of upper bounds that are in interplay with $a_*$. Although by approaching the rotation parameters to the extreme limit, the allowable range of both hair parameters becomes narrower, the hairy Kerr BH solution is a more promising candidate to play the role of the alternative compact object instead of the standard Kerr BH. The lack of tension between hairy Kerr BH with the current observation of the EHT shadow of the M87* SMBH carries this message that there is the possibility of NHT violation.
}

\keywords{Hariy Kerr black hole, Hairy Kerr naked singularity, No-hair theorem, Event horizon telescope}

\begin{document}
	%\date{\today}
	
	\maketitle
	\flushbottom
	
	%%%%%%%%%%%%%%%%%%%%%%%%%%%%%%%%%%%%%%%%%%%%
	\section{Introduction}\label{intro}
	%%%%%%%%%%%%%%%%%%%%%%%%%%%%%%%%%%%%%%%%%%%%
	For a long time there was controversy whether or not the idea of the black hole (BH) solution of the Einstein equation was relevant to real-universe physics and either is a mere fancy idea. Since general theory of relativity (GTR) was developed we have been seen gathering indirect evidence in favor of the existence of BHs, in an extensive range of astrophysical environments and celestial objects. However, we still fail to test the most essential concept required in the BH i.e. event horizon that is no-return membrane to mask the internal properties of BH. There are two well known observational channels that expected to have enough potential to test the geometry of spacetime around the event horizon up to high accuracy: gravitational waves (GWs), and very-long-baseline-interferometry (VLBI) imaging of BH shadows. The former \cite{TheLIGOScientific:2017qsa,Monitor:2017mdv}, has already supplied powerful tests of GTR \cite{Abbott:2018lct} such that refuted or strongly constrained many well-known modified theories of gravity (e.g. see \cite{Baker:2017hug}-\cite{Mastrogiovanni:2020tto}). The latter namely the idea of taking advantage of technology VLBI to image of BH shadow was appeared for the first time in \cite{Falcke:1999pj}. It is indeed the central technology used in the \textit{``Event Horizon Telescope''} (EHT) which is a worldwide network of radio telescopes to record a picture of a dark compact object enclosed by a light ring \cite{Akiyama:2019cqa}-\cite{Akiyama:2019eap}. This picture that released in April 2019, is named \textit{``M87* BH shadow''}  and actually addresses the existence of a supermassive BH (SMBH) deployed in the center of the neighbor supergiant elliptical galaxy Messier 87. This extraordinary picture that without exaggeration is one of the most revolutionary images of physics in the 21st-century disclosed the fact to us that BHs are one of the essential ingredients in the real universe's puzzle. Based on GTR,
	BH shadow \footnote{The idea of the BH shadow actually had born in the seminal papers \cite{Synge:1966okc,Luminet:1979nyg} and \cite{Bardeen:1973} released in the 70s for the  Schwarzschild BH and spinning Kerr, respectively. In the wake of these investigations, it is proven that in the absence of rotation we deal with a perfect circular shadow for the BH, while by considering it the shape of the shadow is elongated as a consequence of the dragging effect.} is made by the null geodesics arising from photons with critical angular momentum in the strong bending region of light ray  \cite{Chandra,Cunha:2018acu} (see also review paper \cite{Perlick:2021aok}). In other words, depending on the angular momentum of photons that follow different pathways, some of them with critical angular momentum, via whirling around the BH, can create the unstable photon sphere indicating the boundary of the shadow.
	This is exactly the same shining halo that we see in the EHT image of M87* BH shadow which is expected to have been built of photons from the hot gas around the BH that have been bent before getting to the telescope. Other sets of photons with small and large angular momentums do not have a contribution to making the BH shadow. Note that BH shadow in comparison with the strong gravitational lensing enjoys this benefit that due to creating two-dimensional dark regions against one-dimensional lensing images, are far easier to be observed \cite{Lima:2021cgb}.
	BH shadow has also the capability that can be utilized in cosmological probes as a standard ruler \cite{Tsupko:2019pzg,Qi:2019zdk} (see also criticisms exerted to this idea in \cite{Vagnozzi:2020quf}).  However, the shadow is not a feature characterizing BHs since also the
	naked singularity or wormholes, as alternative compact objects, can display that. As a consequence, for two reasons it is recommended that we adopt a conservative approach in confronting the EHT image of shadow \cite{Cunha:2018acu}. The first is due to the lack of detecting thermal
	radiation as the primary characteristic of the existence of the event horizon which denotes
	the BHs \cite{Broderick:2009ph,Bambi:2012bh}, while the second is due to the serious challenge concerning the differentiation of a BH shadow from the shadows relevant to the other alternatives compact objects as naked singularity \cite{Ortiz:2015rma,Shaikh:2018lcc,Joshi:2020tlq} (discussions released in \cite{Virbhadra:2002ju,Virbhadra:2007kw} can be enlightening as well) and wormhole \cite{Ohgami:2015nra,Shaikh:2018kfv,Dai:2019mse, Bambi:2021qfo}.

	Information saved in the EHT image of M87* BH shadow due to the new imaging techniques gives us a well-resolution image of the vicinity of BH, meaning that just like quasinormal modes (QNMs) one can use the shadow as a potential probe to control BH parameters related to a wide class of BH solutions. Of course, some may not welcome this statement since by modeling the M87* by the rotating BH geometry, the observation was found to be in good agreement with Kerr BH as predicted in GTR, meaning that no longer the need to consider alternative BH solutions. However, such a view suffers from two challenges. First, the EHT measurements of M87* have a finite resolution and indeed are not conclusive, namely, there still exists some space for alternative models to simulate the BH image within the relevant observational window \cite{Gan:2021xdl}. Second, this is nothing but provide a null-hypothesis test of the GTR predictions, meaning that compatibility between data and theory does not necessarily tell us anything about the accuracy of the hypothesizes used in theory \cite{Psaltis:2014mca,Psaltis:2020lvx}. In this manner, the confrontation of other theories with these measurements is potentially possible. In other words, using the modified metrics, we can calculate how big the shadow should be if an alternative gravity theory is at work. As a result, uncertainties in the measurement of shadow parameters will allow us to impose some novel constraints on the free parameters of alternative gravities that are consistent with observation. So, it is expected that by this way one can refine the road of metric theories of gravity \cite{Younsi:2016azx,Mizuno:2018lxz,DeFalco:2021klh,Kocherlakota:2021dcv}. However, it is also notable that newly in \cite{Glampedakis:2021oie} the authors have
	warned that for probing the physics beyond GTR (and more generally, for extended theories of gravity) by SMBH shadow, it is required that they have dimensionless coupling constants.

	With the advent of EHT, it has been revived a great deal of interest in computing the shadows associated with the extensive class of BH solutions, mostly come from extended theories of gravity, non-linear electrodynamics regimes, new physics and etc \cite{Shaikh:2019fpu}-\cite{Wang:2021irh}. Information extracted from the EHT image of M87* BH shadow can be used to refine the metrics extended due to the surrounding of BHs by dark energy and dark matter \cite{Davoudiasl:2019nlo}-\cite{Lee:2021sws}. The window opened by the EHT let us to a novel probe of some fundamental physics in a strong-gravity regime \cite{Bambi:2019tjh}-\cite{Li:2021mzq}. In a number of interesting works, it is shown that the BH shadow can be investigated in interplay with other relevant ideas such as QNMs, deflection light and quasiperiodic oscillations (QPO) \cite{Jusufi:2019ltj}-\cite{Campos:2021sff}. Additionally,  it is known that the GTR-based magnetohydrodynamic simulations of the magnetized accretion flow onto BH can reveal more details of the BH shadow, see e.g \cite{White:2019wix,Nathanail:2020wap,Bronzwaer:2020vix,Cruz-Osorio:2021gnz}.
	
	One hypothesis that seems to play a fundamental role in disclosing the nature of BH is known as the \textit{``no-hair theorem"} (NHT). It asserts that all astrophysical BHs close to equilibrium state are described by two parameters: mass $M$ and spin $a$, indicating the Kerr metric \cite{Carter:1971zc,Ruffini:1971bza}. Sometimes it also refers to as the Kerr hypothesis which excludes any hairy Kerr solution \cite{Robinson:1975bv}. Note that one cannot merely extract the hair solutions, claiming that the NTH is violated, since the stability status of the hairy solution is of high importance and deserves a deep analysis.  Actually, if it suffers from instability, then the hair parameter will not coexist with the main parameters involved in the metric and thereby NHT remains valid \cite{Rahmani:2020vvv}.
	The significant of NHT is that, unlike what GTR teached us, if BHs have hair, then we would need to change standard GTR \footnote{Despite that yet astrophysical observations have not revealed tension with NHT, there are a number of hints of the theoretical level until observational ones that conduct us to the fact that the present understanding of gravity is faulty and subsequently may the NHT not be a fundamental truth \cite{Cunha:2019ikd} (see also review paper \cite{Capozziello:2011et}).} \cite{Loeb:2013lfa}. More precisely,  the message of NHT is that,  the Kerr metric is the only asymptotically flat stationary, axisymmetric, vacuum solution in GTR with an event horizon which its outside region is free of any pathologies like singularity and closed timelike loops. Subsequently, in any parametric metric that deviate from GTR should not appear at least one of these pathologies. In case of the existence of these issues, the relevant parametric metric is not able to provide a healthy framework in order to probe the validity of NHT.
	Depending on the astrophysical applications, this kind of pathologies may create some troubles.  In tests of the NHT that only involve the orbits of objects at far-away from the
	horizon (as is the case for the motion of stars or pulsars around a BH), the mentioned pathologies have no role. However, they are very vital for the study of accretion flows around BH since the electromagnetic radiation comes predominantly from the close to  event horizon \cite{Johannsen:2011mt}. With this argument, the emission from accretion flows around BHs is most efficient for tests of the NHT (in a strong-gravity regime) with the electromagnetic spectrum X-ray observations  \cite{Johannsen:2010xs,Johannsen:2010ru,Johannsen:2010bi,Bambi:2016sac,Wang-Ji:2018ssh,Abdikamalov:2020oci}. Consequently, to perform a strong-gravity test of the NHT, very accurate modeling of the outside region of the BH spacetime is required.
	To date, there exist different approaches that are able to modeling parametric departures from the standard Kerr metric.
	By admitting this, many efforts have been done in the direction of testing parametric BH metrics include the hair parameter(s) in addition to mass and spin by GW observational channel, see e.g \cite{Glampedakis:2005cf,Li:2007qu,Apostolatos:2009vu,Isi:2019aib,Islam:2019dmk,CalderonBustillo:2020tjf}. In \cite{Cardoso:2016ryw}, one also can be found detailed discussions about how to measure the no-hair properties of BHs via electromagnetic as well as GW window.	The release of EHT image gives the opportunity to hire this approach namely using different parametric frameworks including one or more hair parameters to evaluate NHT in strong-field regime.

	In this direction,  Johannsen \& Psaltis \cite{Johannsen:2011dh}  without restoring to an additional matter field have built a well-behaved hairy Kerr spacetime metric with just one dimensionless                                                                                                                                                                                     hair parameter\footnote{It is notable that there is also other parametric hairy BH solutions with two additional parameters that have been developed in the absence of any fundamental matter field \cite{Ovalle:2020kpd,Contreras:2021yxe}. It newly was subject to the EHT observations of M87* \cite{Afrin:2021imp}. It would be interesting to know that within generic metric theories of gravity also built some parametric frameworks to describe the spacetime of BHs with spherical symmetry \cite{Rezzolla:2014mua} and axisymmetry \cite{Konoplya:2016jvv}, which have some advantages.}. In this paper, we wish to face it to the EHT image recorded of the M87* shadow. Incorporating the hair parameter into the standard Kerr metric riches it for two perspectives. First, depending on the sign of hair parameter, we are dealing with two types of deformed solutions: Kerr naked singularity and Kerr BH. The latter enjoys this benefit that up to the extreme limit of spin everywhere outside of the horizon is free of singularity. Second, it is dimensionless, in agreement with discussions proposed in \cite{Glampedakis:2021oie}.  There it is shown that an essential requirement to probe the possible deviations of the Kerr metric by SMBH shadow is that the non-Kerr character(s) be dimensionless.
	An important point that one should pay attention to about the underlying metric is that it in essence is not a direct vacuum solution of the Einstein equations rather indeed it is derived in a kind of perturbative procedure in order to embed the various possible departures from the Kerr solution in the modified gravities \cite{Atamurotov:2013sca}. In this way, we can evaluate the validity of NHT  via a rich and well-behaved parametric framework deviated from standard Kerr.
	Throughout this paper, motivated by data analysis of M87* by the EHT team \cite{Akiyama:2019fyp}, as well as simulations of the twist of the light emitted from the Einstein ring around M87* shadow \cite{Tamburini:2019vrf}, we will restrict ourselves to high rotation situations particularly within the range $a=(0.9\mp0.1)M$. There is another hint for adopting this strategy and that is the Kerr BH with high rotation is one of the known candidates to explain high energy jets outflow from the center of galaxies as a host of  SMBHs \cite{Blandford:1977ds}. Also, recently in \cite{Ng:2020ruv} shown that the fast-spinning BHs can be utilized in the search for dark matter particles, particularly ultralight bosons such as axions. Actually, it is shown that in the presence of the high rotation BHs the ultralight bosons in some mass ranges are ruled out which is of great value in the search for dark matter.
	
	The outline of the current paper is as follows.
	In Section \ref{frame} we briefly review the hairy Kerr spacetime metric made by Johannsen and Psaltis in \cite{Johannsen:2011dh}.
	There we also provide some discussions and calculations on the essential aspects of its geometry and geodesic. We go to Section \ref{shadow} in order to extracting the shadows related to hairy rotating spacetime at hand. To observational probe of the hair parameter, Section \ref{EHT} is devoted to the strong-field test of NHT through a confrontation between shadows obtained for the underlying hairy Kerr spacetime and EHT measurements of M87* SMBH shadow. To do so, we employ three shadow observables oblateness, deviation from circularity, and shadow diameter.
	We close this paper in Section \ref{final} by supplying a highlighted note on the analyzes along with acquired results.
	For simplicity, we adopt the Planck units $c=G_N=\hbar=1$, throughout this paper.
	
	%%%%%%%%%%%%%%%%%%%%%%%%%%%%%%%%%%%%%%%%%%%%%%
	\section{A framework suitable for strong field testing of NHT}
	\label{frame}
\begin{figure}
		\begin{center}	
			\includegraphics[width=0.7\linewidth]{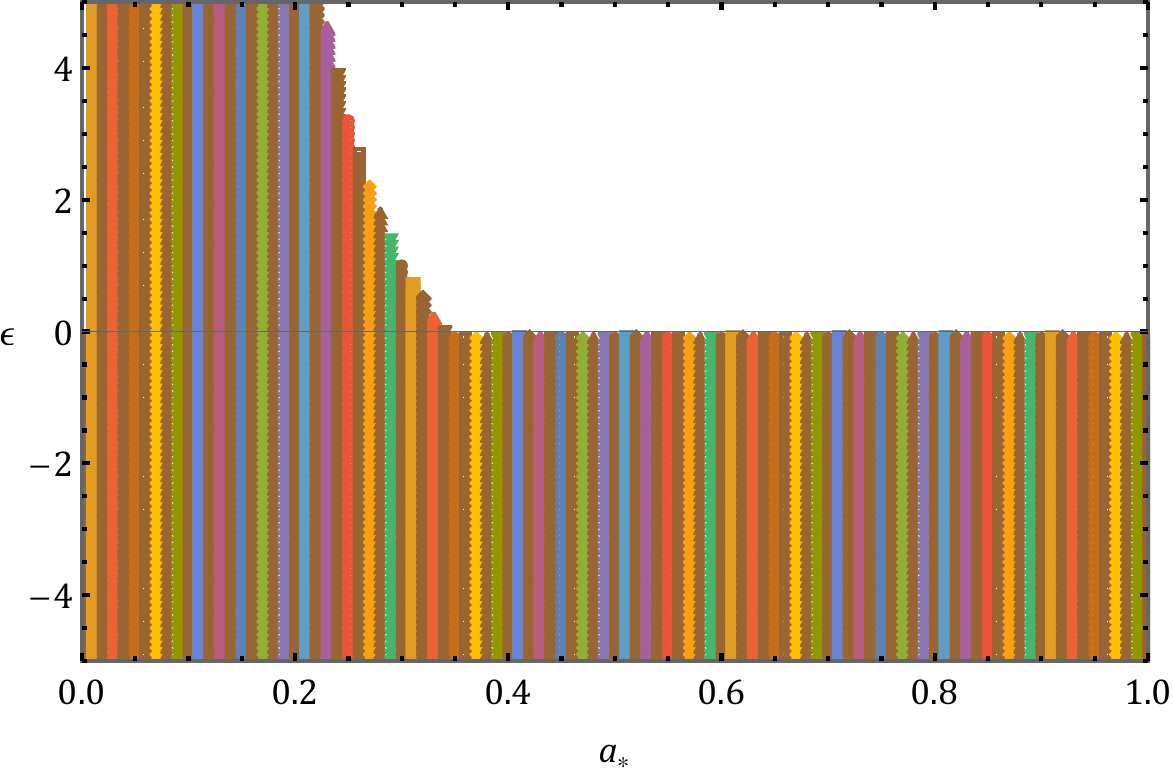}
			\caption{The colored and colorless regions in the $a_*-\epsilon$ plane show the existence of BH solution and naked singularity in deformed Kerr spacetime described by the metric (\ref{metric}), respectively.}
			\label{BHcond}
		\end{center}
	\end{figure}

	%%%%%%%%%%%%%%%%%%%%%%%%%%%%%%%%%%%%%%%%%%%%%%%%%%%%%%%%%%%%%%%%
	The framework under our attention to probe the signatures of possible NHT violation is a deformed Kerr-like metric as follows \cite{Johannsen:2011dh}
	\begin{align}
		ds^2 =&g_{\mu\nu}dx^{\mu}dx^{\nu}=
		-\bigg(1-\frac{2Mr}{\rho^2} \bigg) (1+h) dt^2 + \frac{\rho^2 (1+h)}{\Delta + a^2 h \sin^2\theta} dr^2 + \rho^2 d\theta^2 -\\
		&
		- \frac{4Mar \sin^2\theta}{\rho^2} (1+h)  dt d\phi +  \bigg[\rho^2 \sin^2\theta+ a^2\sin^4\theta \bigg(1+\frac{2Mr}{\rho^2} \bigg) (1+h)\bigg]  d\phi^2\,, \nonumber \\
		&\rho^2 = r^2 + a^2 \cos^2\theta,~~~~~~~\bigtriangleup = r^2-2Mr+a^2,
		\label{metric}
	\end{align}
	where represents a stationary axisymmetric, and asymptotically flat spacetime in the standard Boyer-Lindquist coordinates $(t,r,\theta,\phi)$. In addition to mass $M$ and spin $a$, the underlying spacetime metric deals with some other parameters involved in the function $h(r)$. The general form of the function $h(r)$
	reads as
	\begin{eqnarray}\label{h}
		h(r)=h =\sum_{n=0}^{\infty}\epsilon_n \bigg(\frac{M}{r}\bigg)^n~,
	\end{eqnarray} where indeed $\epsilon_n$ are deformation parameters that measure departure from standard Kerr metric. In order to identify the leading term of  $\epsilon_n$, one have to consider some requirements. First, in limit $r\longrightarrow \infty$, the metric
	(\ref{metric}), can be re-expressed as
	\begin{align}
		ds^2 \approx &-\left(1 - \frac{2M}{r} + h(r) \right) dt^2
		- \frac{4a(1+h(r))}{r}\sin^2\theta dtd\phi + \left( 1 + \frac{2M}{r} + h(r) \right) dr^2 + \nonumber \\
		& + r^2 (d\theta^2 + \sin^2\theta d\phi^2)~,
		\label{asymptmetric}
	\end{align}
	where in the case of $\epsilon_0=0=\epsilon_1$ in (\ref{h}), the asymptotic flatness of the metric, is guaranteed. Second, comparing with the parameterized post-Newtonian (PPN) approach \cite{Will:2005va}
	thereby, the asymptotic form of spacetime reads as
	\begin{align}
		ds^2 = &-\left( 1-\frac{2M}{r} +2(\beta-\gamma)\big(\frac{M}{r}\big)^2\right) dt^2 +\bigg(1+2\gamma\frac{M}{r}\bigg) dr^2 + r^2 d\theta^2+ \sin^2\theta d\phi^2 ,
		\label{asympt}
	\end{align}
	where $\beta$ and $\gamma$ denote the dimensionless PPN parameters. By comparing the constraint on $\beta$ (i.e. $\beta-1\leq2.3\times 10^{-4}$ \cite{Williams:2004qba}) and the asymptotic metric (\ref{asymptmetric}), one finds that vanishing of the parameter $\epsilon_2$ is essential. As a result, these requirements lead to the discarding terms in Eq.  (\ref{h}) up to the second power of $M/r$, meaning that the leading deformation parameter $\epsilon_n$ comes with the third power of $M/r$
	\begin{equation}\label{hf}
		h(r,\theta)= \epsilon_3 \frac{M^3 r}{\rho^4} ~.
	\end{equation}
	For simplicity, after now on we drop index 3 and use the label $\epsilon$ for the deformation parameter which is also called the hair parameter. Clearly, if $\epsilon=0$,
	the typical Kerr BH is restored. Although the possible value of the deformation parameter $\epsilon$ can be positive as well as negative, for the former case we will not have a Kerr like BH solution but instead, gives us a Kerr naked singularity. This statement can be verified by calculating the location of the event horizon of the underlying metric via finding the roots of the equation $g_{t\phi}^2-g_{tt}g_{\phi\phi}=0$, see Fig. \ref{BHcond}. It is quite clear that for case of high rotation (as our main interest) $\epsilon>0$ address deformed Kerr naked singularity while $\epsilon<0$ denotes deformed Kerr BH solution. Given the fact that the shadow is not the mere feature of BHs, so throughout this paper, we will subject to test both as alternative options for the standard Kerr BH. Concerning the deformed Kerr BH solution of the metric (\ref{metric}), it is shown that regions outside the event horizon enjoy the regularity and lack of singularity up to maximum values of the dimensionless spin parameter $a_*=a/M$ \cite{Johannsen:2011dh}. This represents a Kerr-like BH solution well-behaved in everywhere regions outside of the event horizon.
	
	The mention of two points here about $h(r,\theta)$ is worthy.
	First, chosen $h(r,\theta)$ in Eq.  (\ref{hf}) does not damage the stationary and axisymmetric properties of Kerr BH. Second, in the presence of $h(r,\theta)$,
	the Einstein tensor of spacetime metric (\ref{metric}) is no longer zero. Namely, one has to consider the underlying spacetime metric as a vacuum spacetime of set of field equations modified relative to GTR.
	In other words, the metric (\ref{metric}) is extractable via a kind of perturbative manner in which one can in the Kerr solution embed the varieties of possible deviations expected from modified gravity theories. 
	
	\subsection{Geodesic equations }
	In the direction of our aim, now we evaluate the evolution of photon's pathway around metric spacetime
	(\ref{metric}). To do that, as usual, one should be study the geodesics in the underlying spacetime, via the following Hamilton-Jacobi equation
	\begin{equation}
		\frac{\delta S}{\delta \tau} = -\frac{1}{2} g^{\mu \nu}\frac{\delta S}{\delta x^{\mu}} \frac{\delta S}{\delta x^{\nu}} \label{eq1},
	\end{equation}
	where $\tau$ and $S$ denote an affine parameter  and Jacobi action, respectively. Because of we are deal with the separable solution of the above differential equation, thereby, the Jacobi action in case of fixing photon as test particle with zero rest mass, reads as follows
	\begin{equation}
		S = - Et + L \phi + S_{r}(r) + S_{\theta}(\theta) \label{eq2},
	\end{equation}
	where $E$ and $L$ are respectively the conserved energy and angular momentum that are connected with the killing vector fields $\partial_t$ and $\partial_\phi$. By putting Eq.~(\ref{eq2}) into (\ref{eq1}) along with using the inverse elements of the metric (\ref{metric}) after doing some algebra, we arrive at
	\begin{align}
		& \frac{\delta S_{r}}{\delta r} = \frac{1 +h(r,\theta)}{\Delta + h(r,\theta) a^2 \sin^2{\theta}}\sqrt{\mathcal{R}(r)}~, \label{eqr}\\
		&\frac{\delta S_{\theta}}{\delta \theta} = \sqrt{\varTheta(\theta)}\label{eqteta}\,\, ,
	\end{align}
	where
	\begin{eqnarray}
		\mathcal{R}(r)&=&\bigg((r^2 + a^2)E - a L \bigg)^2-\Delta \bigg(\mathcal{Q} +(L - a E)^2 \bigg)~,\label{R}\\
		\varTheta(\theta)&=&\mathcal{Q}-L^2 \cot^2{\theta} +a^2 E^2\cos^2{\theta}  \label{teta}\,\, ,
	\end{eqnarray} Here $\mathcal{Q}$ is so called Carter constant as the third conserved quantity in addition to $E$ and $L$ which play the role of a separation constant. Finally, some algebra gives the following four equations of motion
	\cite{Bambhaniya:2021ybs}
	\begin{align}
		&	\rho^2\frac{dt}{d\tau}=\frac{r^2+a^2}{\Delta + h(r,\theta) a^2 \sin^2{\theta}}\bigg((r^2 + a^2)E - a L\bigg)+
		\frac{\Delta a}{\Delta + h(r,\theta) a^2 \sin^2{\theta}}\bigg(L- a E\sin^2{\theta}\bigg)~,\label{m1}\\
		&	\rho^2\frac{dr}{d\tau}=\sqrt{\mathcal{R}(r)}~, \label{m2}\\
		&\rho^2\frac{d\theta}{d\tau}=\sqrt{\varTheta(\theta)}~,\label{m3}\\
		&	\rho^2\frac{d\phi}{d\tau}=\frac{a}{\Delta + h(r,\theta) a^2 \sin^2{\theta}}\bigg((r^2 + a^2)E - a L\bigg)+
		\frac{\Delta}{\Delta \sin^2{\theta} + h(r,\theta) a^2 \sin^4{\theta}}\bigg(L- a E\sin^2{\theta}\bigg)~,\label{m4}
	\end{align}
	which identify the trajectory of photon (null geodesics) around deformed Kerr spacetime (\ref{metric}).
	It can be easily seen that by relaxing deformation parameter $\epsilon$ in the above set of equations, these can be written in their standard form. Conventionally, the light ray is identified by impact parameters
	$\xi=\frac{L}{E}$ and $\eta=\dfrac{\mathcal{Q}}{E^2}$ which are defined in terms of three conserved quantities $E,~L$ and $\mathcal{Q}$. By taking these two impact parameters into account thereby, the radial dependence of the effective potential of null geodesics re-express as follows
	\begin{equation}
		\dfrac{\mathcal{R}(r)}{E^2} = \bigg((r^2 + a^2) - a \xi \bigg)^2 - \Delta \bigg(\eta + (\xi - a)^2 \bigg)\,\, .
		\label{rpoten}
	\end{equation}
	As a result, the radial motion of photon obey from following equation
	\begin{equation}
		\big(\frac{dr}{d\tau}\big)^2+U_{eff}(r)=0,~~~U_{eff}(r)=-\frac{\mathcal{R}(r)}{r^4 E^2}~,
		\label{potential}
	\end{equation}
	Applying conditions $\mathcal{R}(r) = 0=\frac{d \mathcal{R}(r)}{dr}$ leads to determine the $r$-constant orbits around the spacetime metric at hand. By these orbits, one can identify the boundary of shadow with the following two impact parameters
	\begin{equation}\label{IP}
		\xi = \frac{r^2(r - 3M) + a^2 (r + M)}{a (M - r)}~,~~~~~~~~
		\eta = \frac{r^3\big(4Ma^2-r(r-3M)^2\big)}{a^2(M - r)^2}~.
	\end{equation}
	Note that for forming the shadow the existence of unstable photon sphere orbits that come from condition $\frac{d^2U_{eff}(r)}{dr^2}<0$ or $\frac{d^2\mathcal{R}(r)}{dr^2}>0$, is essential. As usual, by adopting equatorial orbits $(\theta=\pi/2)$, the radius of the unstable photon sphere orbit is characterized by $\eta=0$. Here one will get nothing but what expected for the standard Kerr i.e. $r_{ph^{\mp}}=2M\bigg(1+\cos\big(\frac{2}{3}\arccos(\mp \mid a_*\mid)\big)\bigg)$ which negative and positive signs denote
	a prograde orbit moving in the same direction as the
	BH rotation, and a retrograde orbit moving against the BH
	rotation, respectively.
	
	The above result, i.e. the expressions in Eq.~(\ref{IP}) plus $\mathcal{R}(r)$ in Eq.~(\ref{rpoten}), all explicitly tell us that
	deformation parameter $\epsilon$ do not affect the radial part of effective potential and the location of unstable photon sphere, as well. As a result, departure from the standard Kerr spacetime via $U_{eff}(r)$ is not distinguishable since the nature of $\mathcal{R}(r)$ in both Kerr and deformed Kerr spacetime, is identical. However, the story does not end here, and in what follows will be showed that the shape of shadow in the map of space is openly under the effect of the deformation parameter $\epsilon$. This can be a smoked gun in order to evaluate the validity of NHT via EHT measurements of M87* SMBH.
	\begin{figure}
		\begin{center}	
			\includegraphics[width=0.8\linewidth]{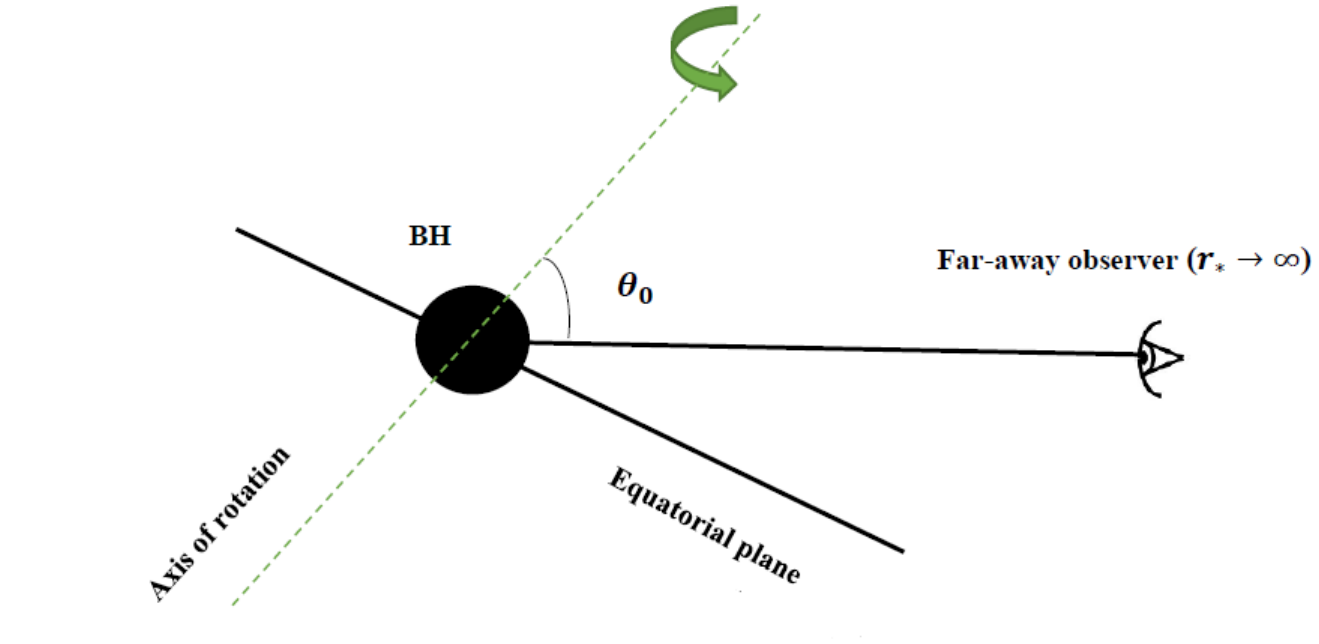}
			\caption{A simple schema of the position of a far-away observer$(r_*\longrightarrow\infty)$ which his/her sight-line with rotating axis of BH has an inclination angle $\theta_ 0$. Note that the inclination angle $\theta_ 0$ is also interpreted as the observation angle.}
			\label{Observer}
		\end{center}
	\end{figure}
	
	\begin{figure}
		\begin{center}	
			\includegraphics[width=0.5\linewidth]{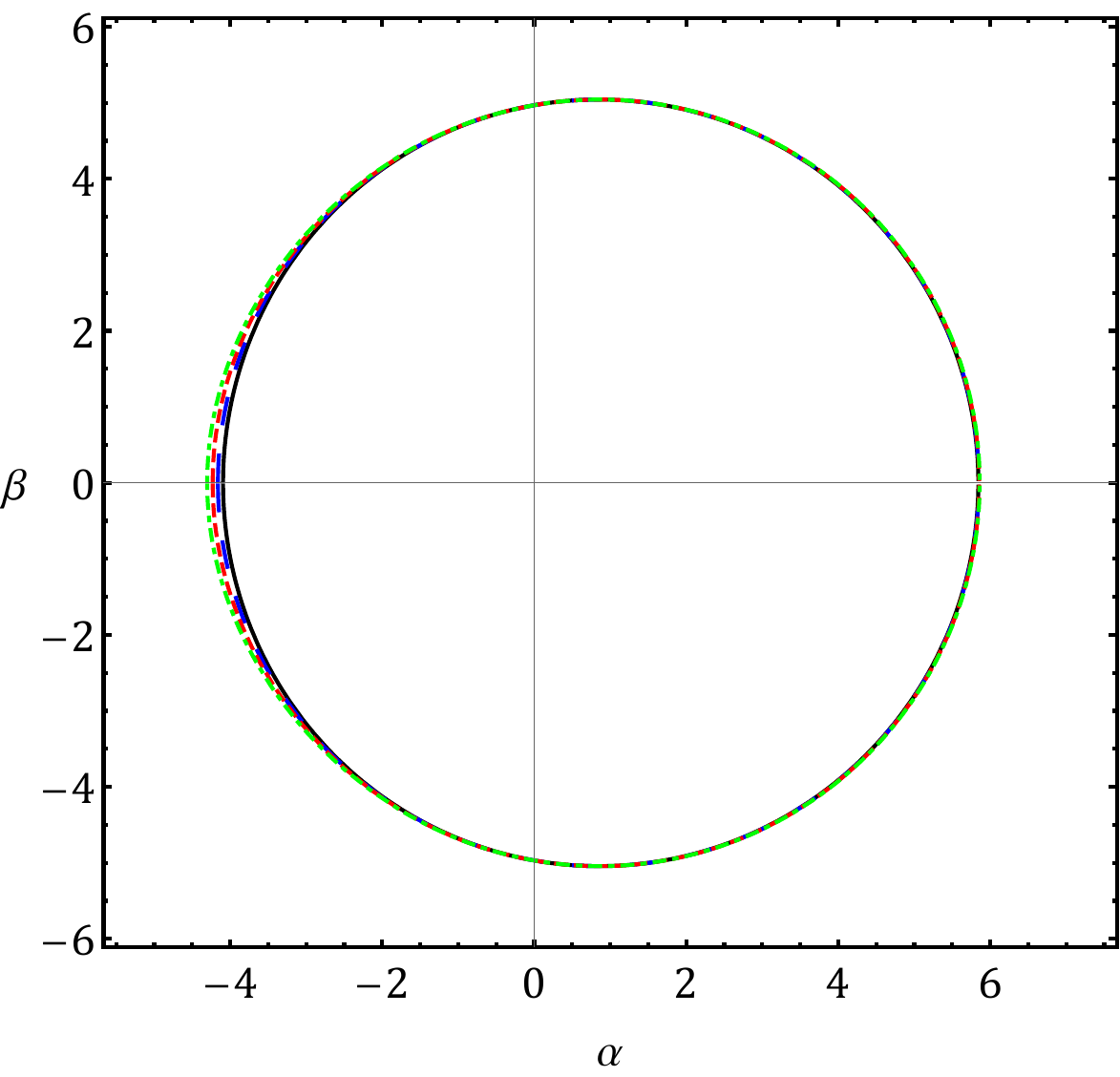}~~
			\includegraphics[width=0.5\linewidth]{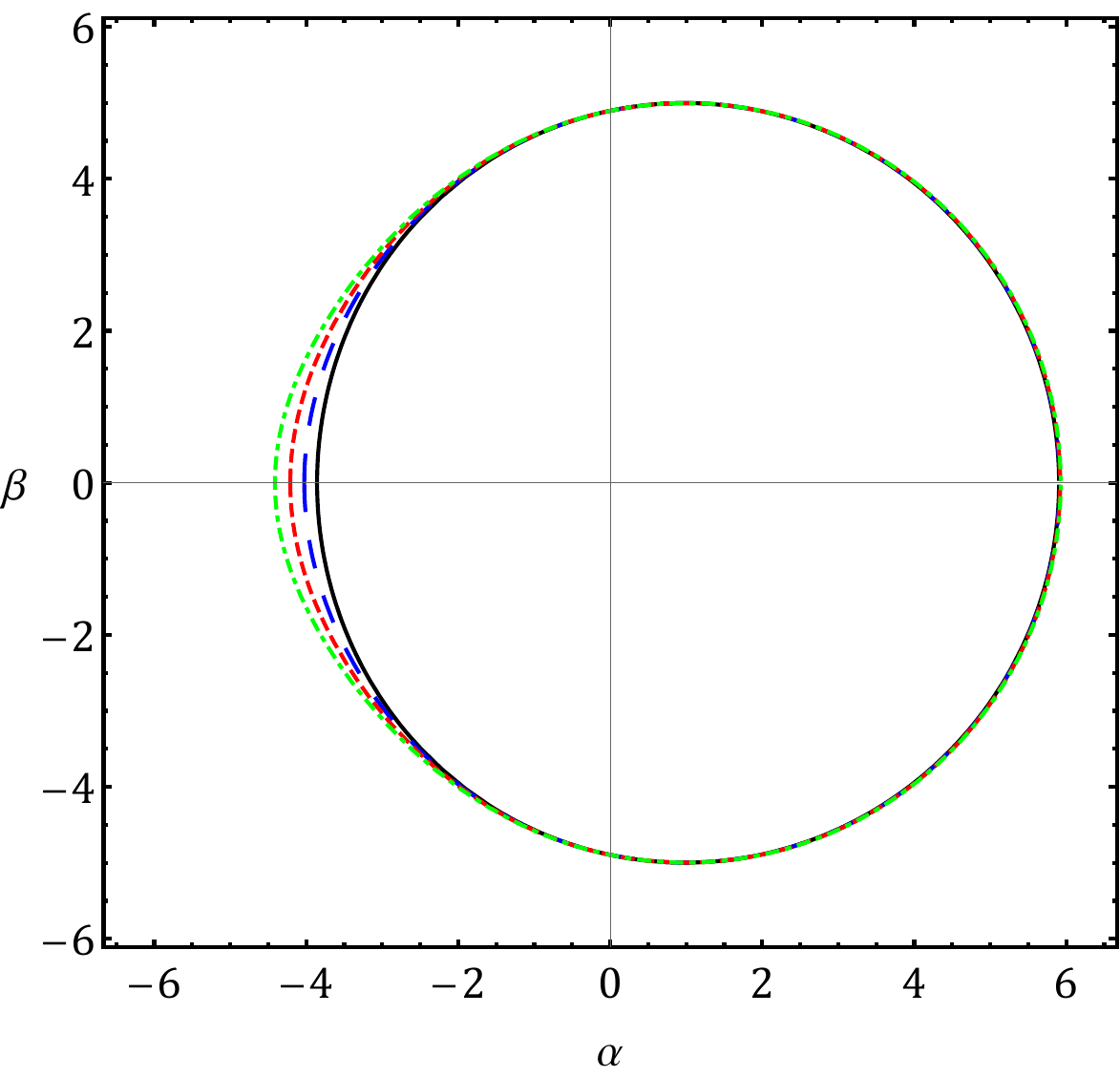}\\
			\includegraphics[width=0.5\linewidth]{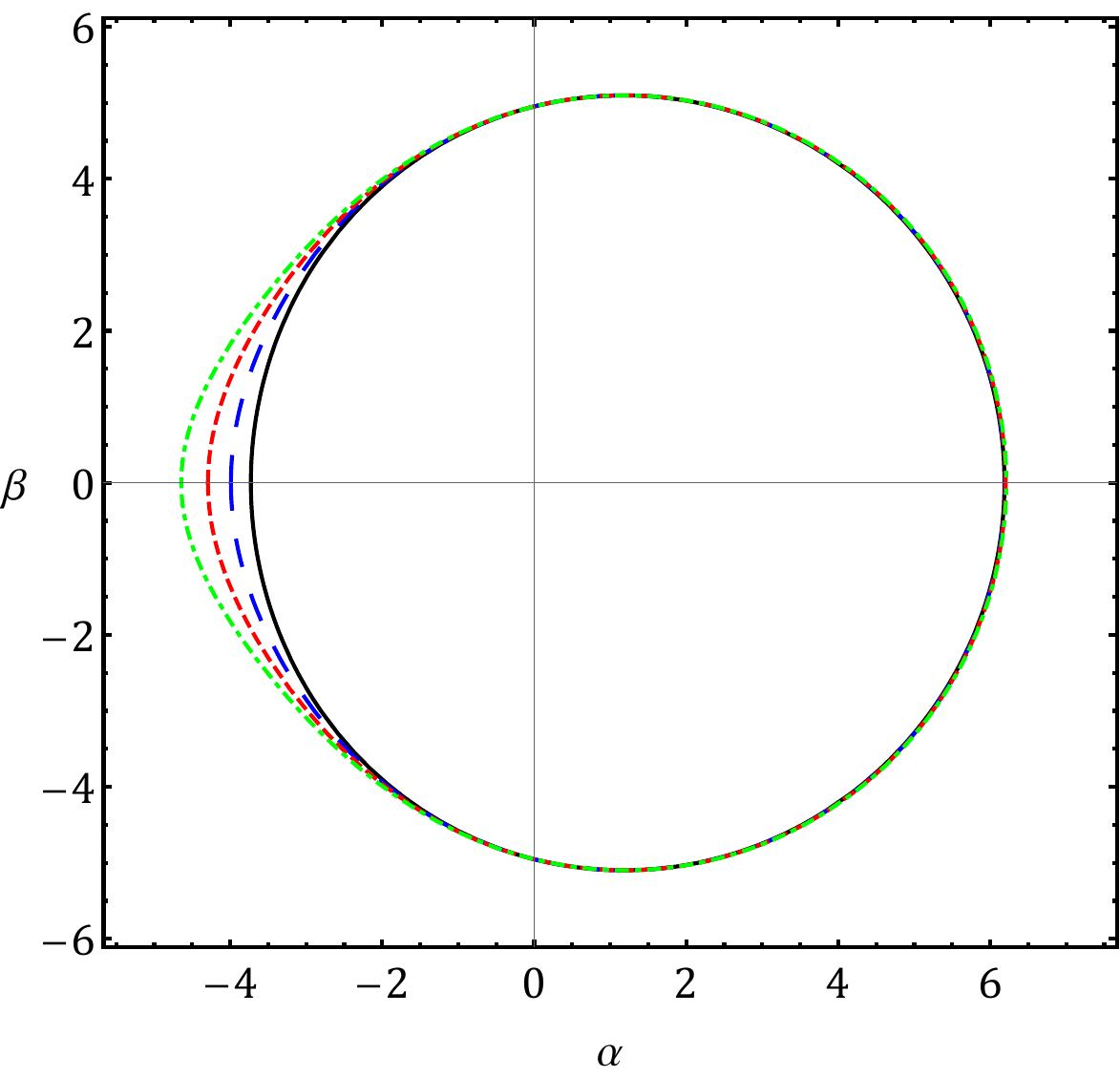}~~
			\includegraphics[width=0.6\linewidth]{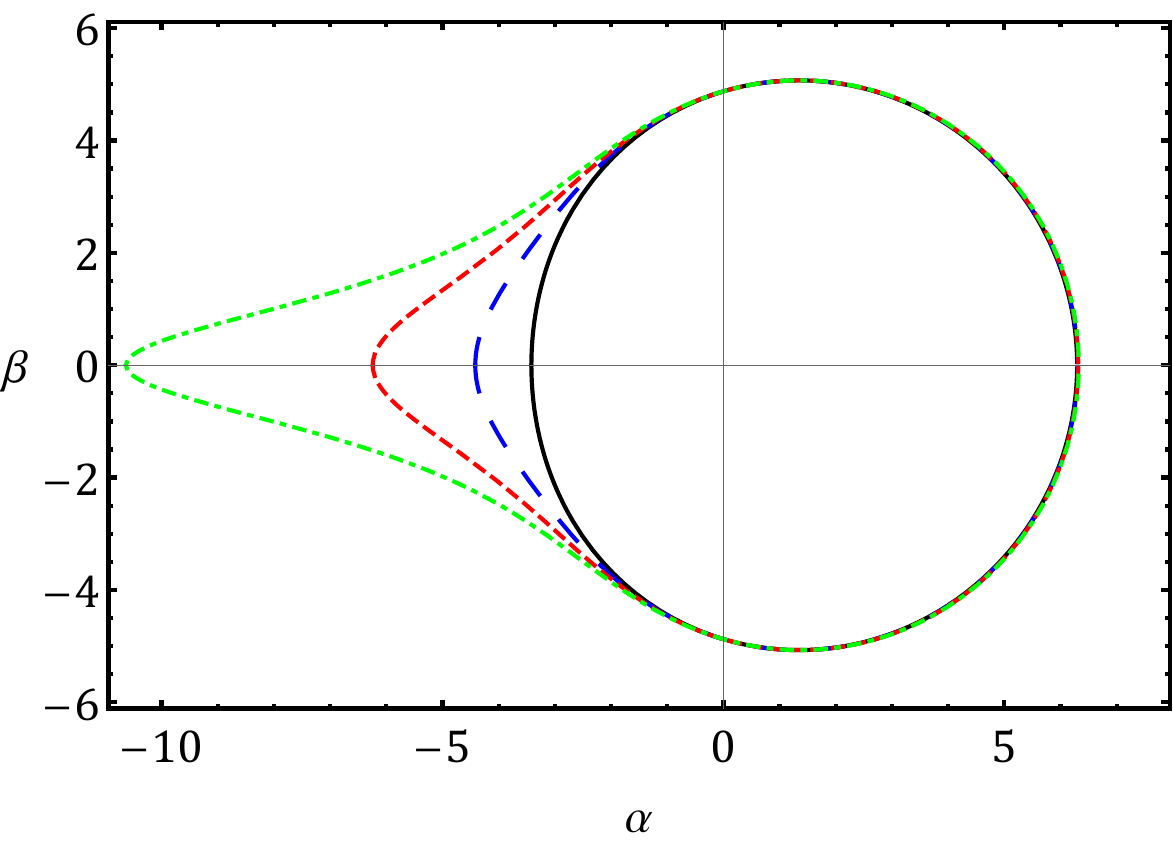}
			\caption{Shadow shapes expected from the Hairy Kerr spacetime (\ref{metric}) for different negative values of $\epsilon:\{0,-1,-2,-3\}$ (from the black-solid curve to green-dot-dashed one) in two different inclination angles $\theta_0$: $30^{\circ}$ (up raw) and $45^{\circ}$(bottom raw). We fixed values $a_*=0.8,~0.9$ in the left and right panels of any raw, respectively.}
			\label{Sh1}
		\end{center}
	\end{figure}
	
	\begin{figure}
		\begin{center}		
			\includegraphics[width=0.5\linewidth]{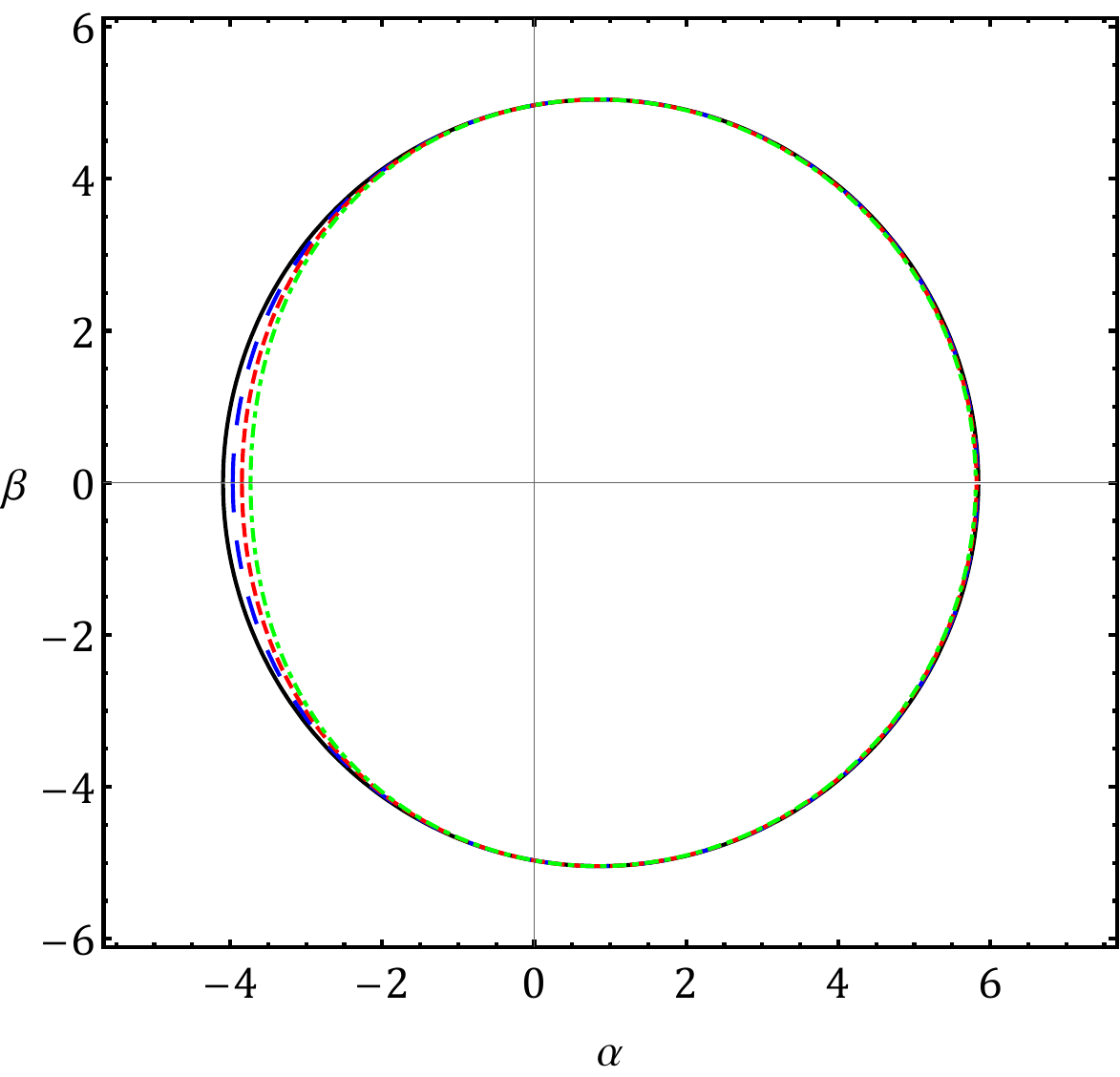}~~
			\includegraphics[width=0.5\linewidth]{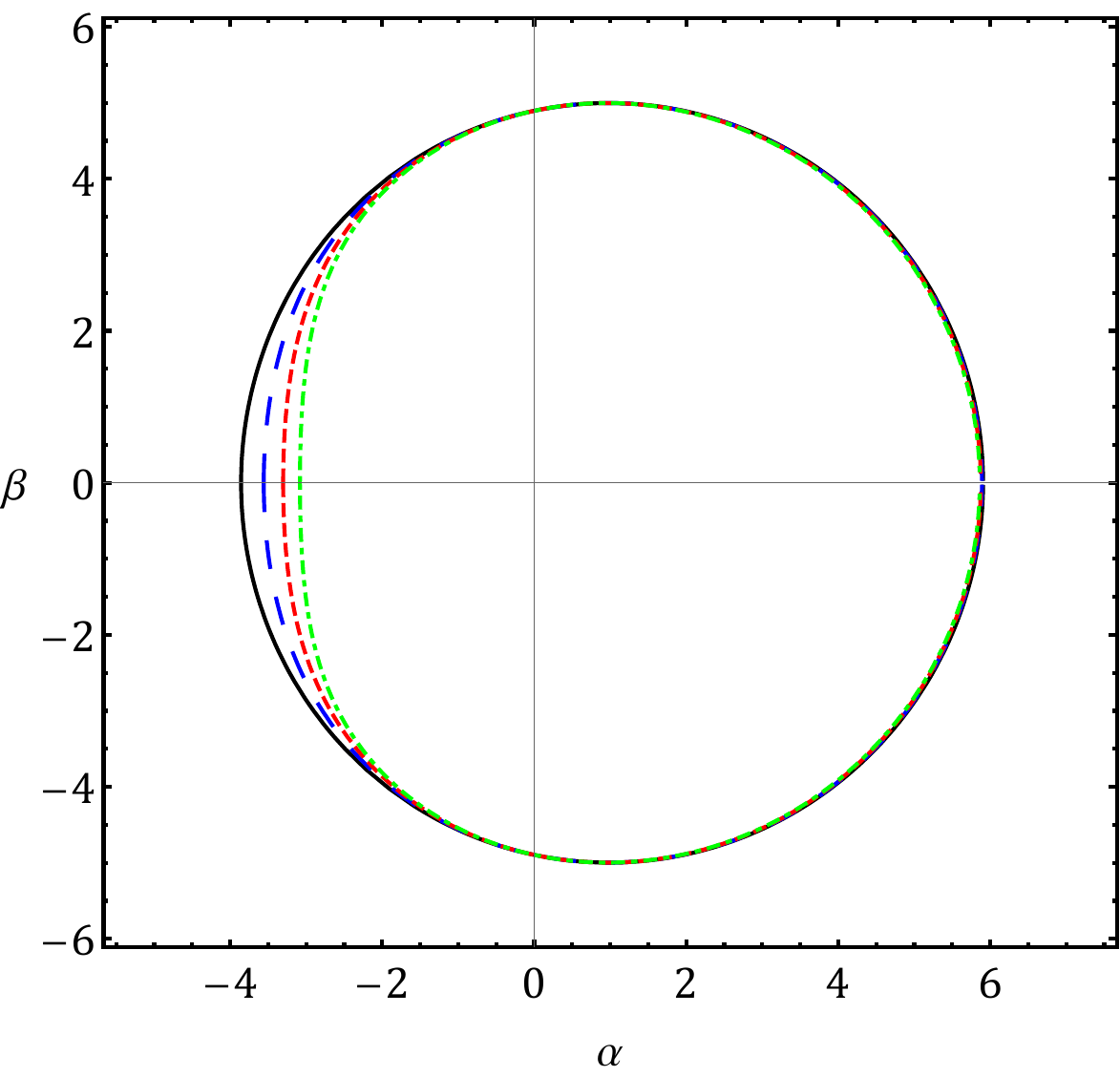}\\
			\includegraphics[width=0.5\linewidth]{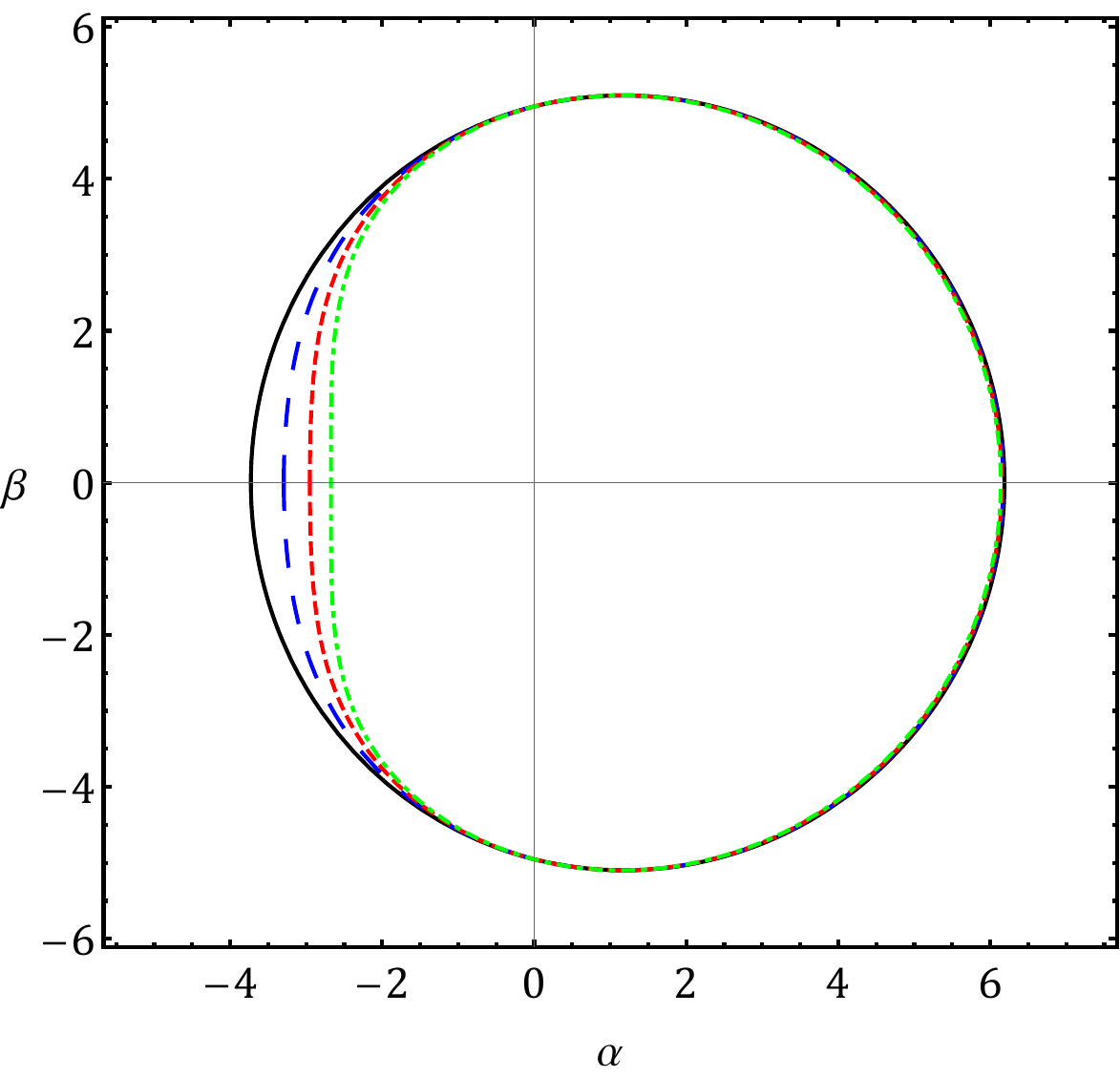}~~
			\includegraphics[width=0.5\linewidth]{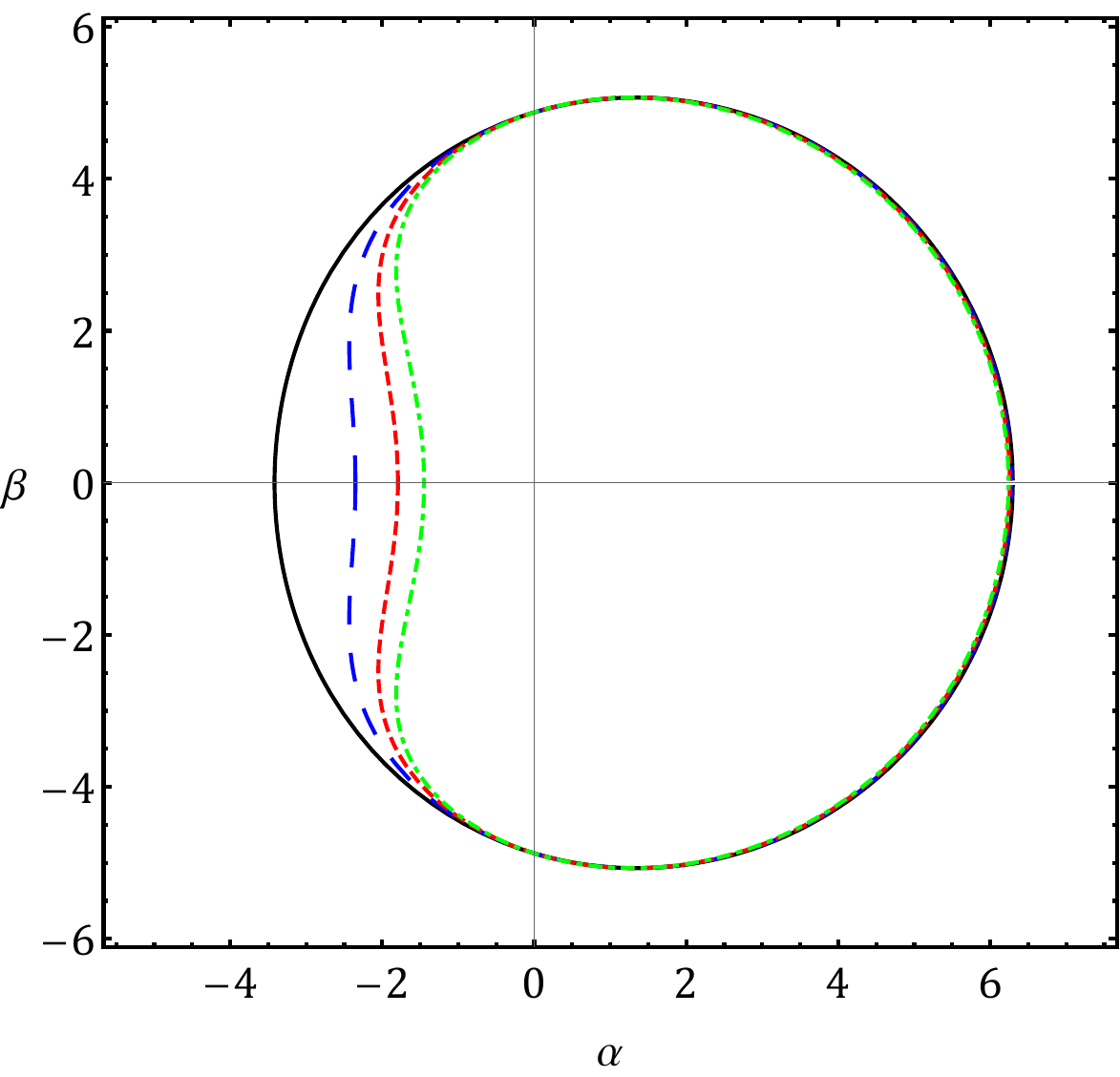}
			\caption{Shadow shapes expected from the Hairy Kerr spacetime (\ref{metric}) for different positive values of $\epsilon:\{0,2,4,6\}$ from the black-solid curve to green-dashed-dotted one. For $\theta_0$ and $a_*$ we set the numerical values used in Fig. \ref{Sh1} with the same arrange.}
			\label{Sh2}
		\end{center}	
	\end{figure}
	
	%%%%%%%%%%%%%%%%%%%%%%%%%%%%%%%%%%%%%%%%%%%%%%%%%%%%%%%%%%%%%%%%%%%
	\section{Shadow of Hairy rotating spacetime}\label{shadow}
	%%%%%%%%%%%%%%%%%%%%%%%%%%%%%%%%%%%%%%%%%%%%%%%%%%%%%%%%%%%%%%%
	If between an observer and a source of light there is a compact object such as BH (or any other compact objects), then the light comes to the observer after a deflection due to the strong gravitational field. However, a part of the deflected light with small impact parameters can be absorbed by the central compact object that leads to a dark shape in the map of the space called the shadow (see, e.g., \cite{Cunha:2018acu}).
	Generally speaking, the shadow is the border area between the photon orbits which is swallowed and scattered by a compact object.
	
	To extract the shadow shape in the underlying spacetime, as can be seen by a faraway observer, one has to derive the expressions of the celestial coordinates $\alpha$ and $\beta$, as \cite{Chandra}
	\begin{eqnarray}
		\alpha = \lim_{r_*\to\infty} \left(-r_*^2 \sin{\theta_{0}} \frac{d \phi}{dr}\right)\,,~~~~~~~~~~~
		\beta = \lim_{r_*\to\infty} \left(r_*^2 \frac{d \theta}{dr}\right)\,,
	\end{eqnarray}
	where $r_*$ and $\theta_0$ are, respectively, the distance of the observer from BH and the inclination angle between the observer line of sight and the rotational axis of the BH, as schematically depicted in Fig.  \ref{Observer}. The celestial coordinates  $(\alpha,\beta)$ actually are the apparent perpendicular distances of the image as observed from the axis of symmetry as well as from its projection on the equatorial plane, respectively. By computing $d\phi/dr$ and $d\theta/dr$ using expressions written in Eqs.~(\ref{m1})-(\ref{m4}) and also taking the coordinate’s limit of a faraway observer $(r_*\longrightarrow\infty)$, then the celestial coordinates functions
	for the deformed Kerr spacetime described by the metric (\ref{metric}), reads as
	\cite{Bambhaniya:2021ybs,Atamurotov:2013sca}
	\begin{eqnarray}\label{cord}
		\alpha = -  \frac{\Delta \xi}{(\Delta + a^2 h \sin^2{\theta_{0}})\sin{\theta_{0}}}\,,~~~~~~~~~
		\beta= \sqrt{\eta + a^2 \cos^2{\theta_{0}} - \xi^2 \csc^2{\theta_{0}}\,\,}~.
	\end{eqnarray}
	The above expressions include this interesting message that though the behavior of $R(r)$ in both Kerr and deformed Kerr spacetime is similar, the shadow shape of deformed Kerr spacetime due to explicitly dependency to the deformation parameter $\epsilon$ is distinguishable from its standard counterpart. It can be a motivation to confirm or refute NHT via confrontation between the hairy Kerr spacetime (\ref{metric}) and some observables related to EHT data. The same thing we will do in the next section.
	
	In Figs.~\ref{Sh1} and \ref{Sh2}, we draw the shapes of the shadows expected from the deformed Kerr spacetime (\ref{metric}) and also its standard counterpart for various values of $\epsilon,~a$ and $\theta_0$. Overall, one can see from these figures that the hairy Kerr spacetime with deformation parameters $\epsilon<0$ and $\epsilon>0$ are more oblate and prolate relative to Kerr $(\epsilon=0)$. However, how more oblate and prolate than Kerr, highly dependent on values of spin parameter $a$, inclination angle $\theta_0$, in addition of values of deformation factor $\epsilon$. More precisely, in a fixed value of $\epsilon$, the deviation of the deformed Kerr spacetime (\ref{metric}) from the standard Kerr becomes more significant, as the values of $a$ and $\theta_0$, increase.
	\begin{figure}
		\begin{center}		
			\includegraphics[width=0.55\linewidth]{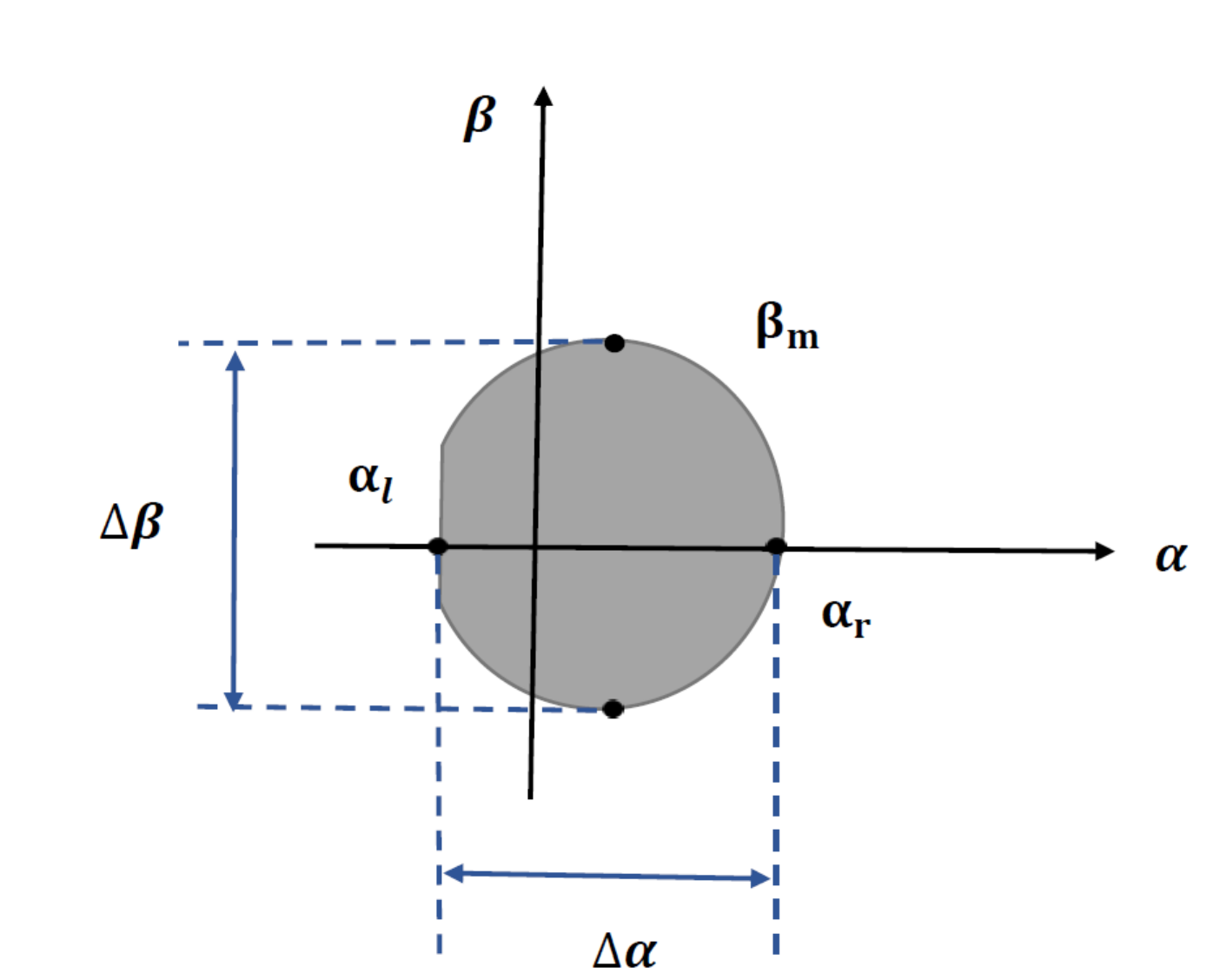}~~~
			\includegraphics[width=0.55\linewidth]{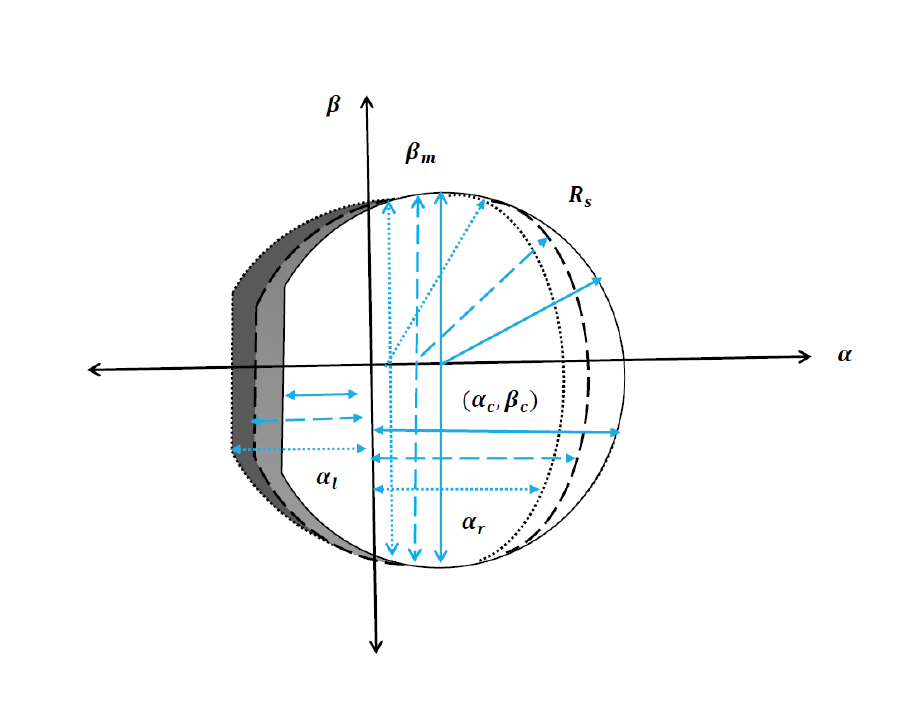}
			\caption{\textbf{Left:} A schematic display of the non-spherical BH shadow. The distances $\alpha_r$ and $\alpha_l$ denote the celestial coordinate $\alpha$ of the most positive and the most negative values supposed by it coordinate. $\pm \beta_m$ are extreme points in the $\beta$-axis. \textbf{Right:} Geometric point $(\alpha_c,\beta_c)$ is build to define the radius of shadow $R_s$ which utilized in computation of DC. Here we displayed three shadows in a frame that have different geometric points $(\alpha_c,\beta_c)$ and radius $R_s$ with the same oblateness $D$. Changes $\bigtriangleup \alpha$ and $\bigtriangleup \beta$ in any three shadows are equal.
				 }
			\label{Shematic}
		\end{center}	
	\end{figure}
	
	\begin{figure}
		\begin{center}		
			\includegraphics[width=0.5\linewidth]{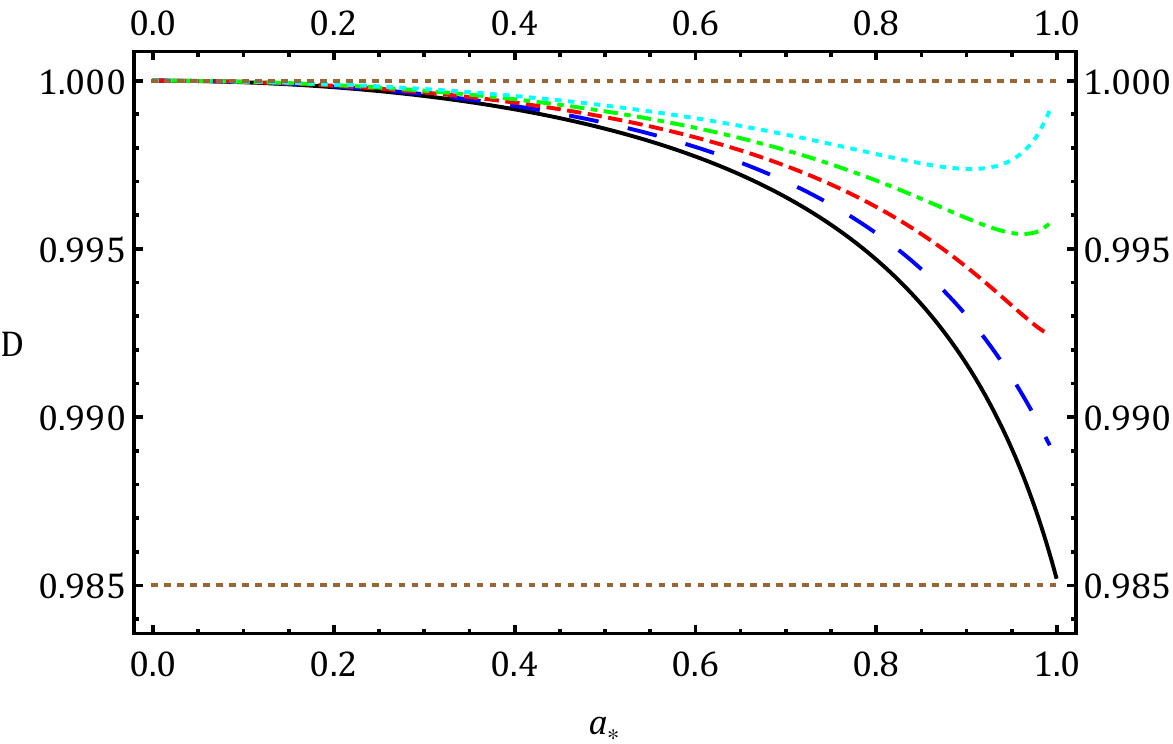}~~
			\includegraphics[width=0.5\linewidth]{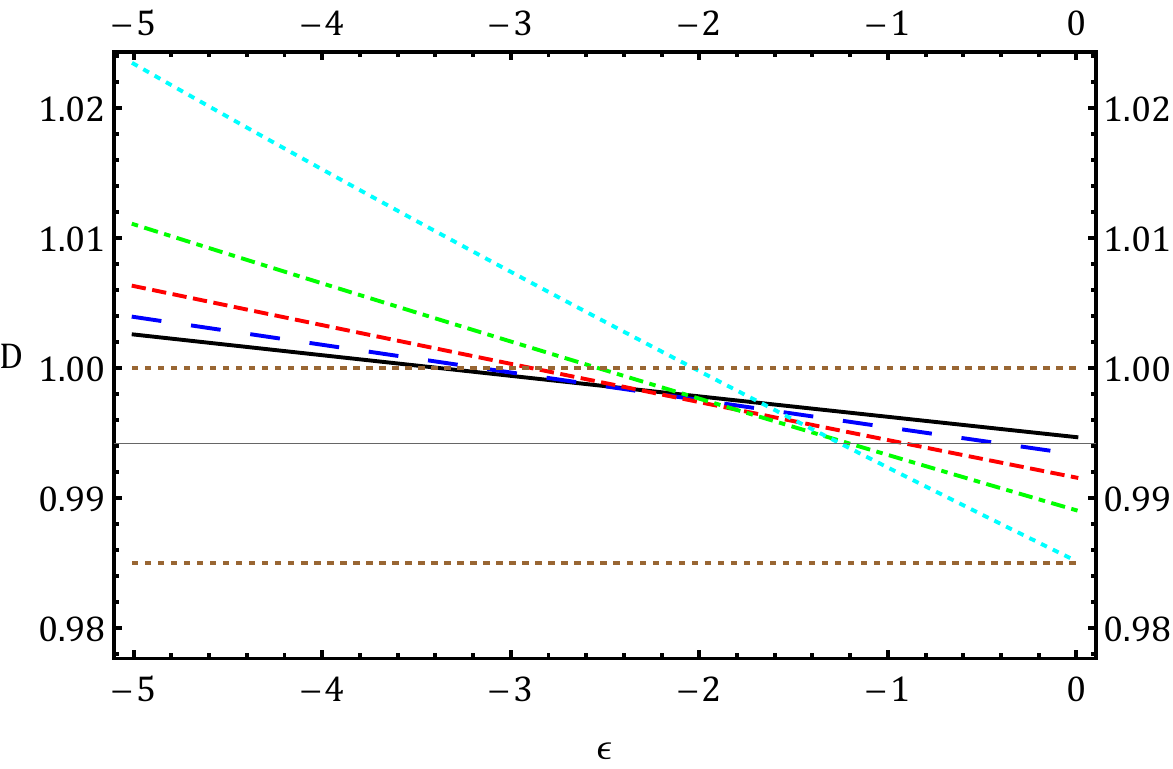}\\
			\caption{\textbf{Left:} Oblateness $D$ in terms of $a_*$ for different positive values of deformation parameter $\epsilon:\{0, -0.5,-1,-1.5,-2\}$ from the black-solid curve to cyan-dotted one, respectively. \textbf{Right:} Oblateness $D$ in terms of $\epsilon$ for several values of $a_*:\{0.8, 0.85,0.9,0.95,1\}$ from the black-solid curve to cyan-dotted one, respectively.
				Here as well as other plots in this section to match the EHT data we fix  inclination angle $\theta_0$=$17^{\circ}$.  Brown-dotted lines denote lower and upper bounds expected from the oblateness which makes an allowed region for its change from the view of the observer with sight-angle  $\theta_0$=$17^{\circ}$.}
			\label{Oblatn}
		\end{center}	
	\end{figure}
	
	\begin{figure}
		\begin{center}	
			\includegraphics[width=0.5\linewidth]{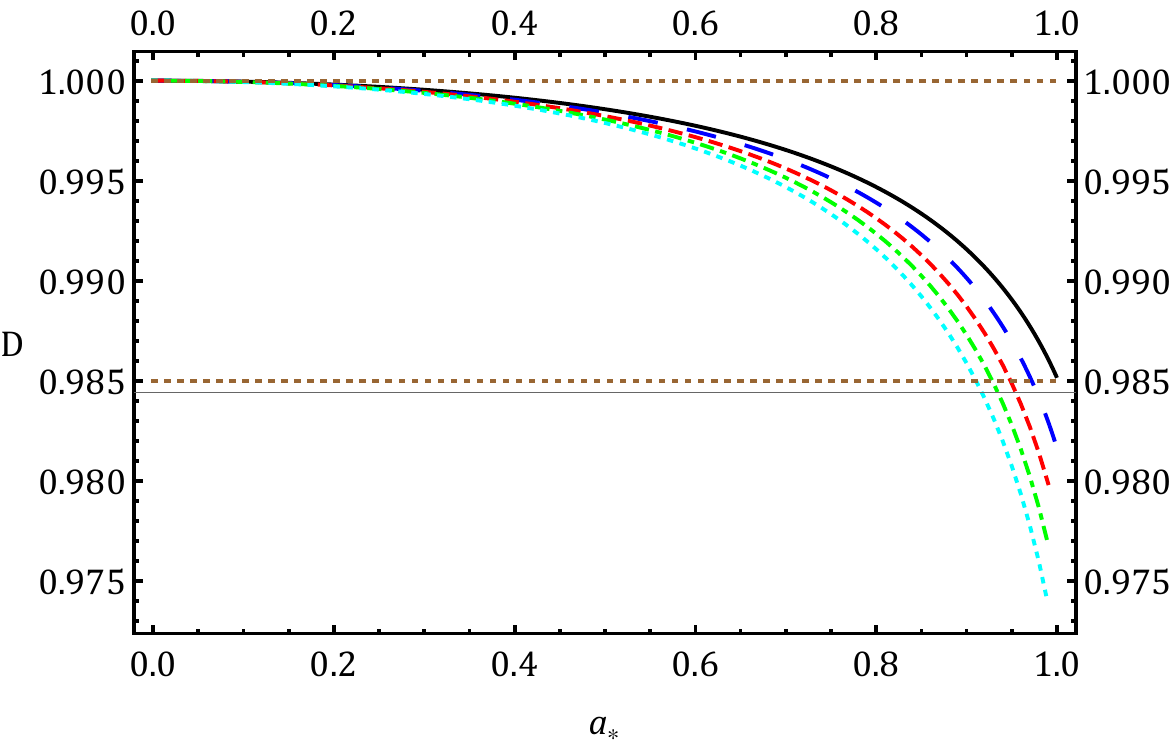}~~
			\includegraphics[width=0.5\linewidth]{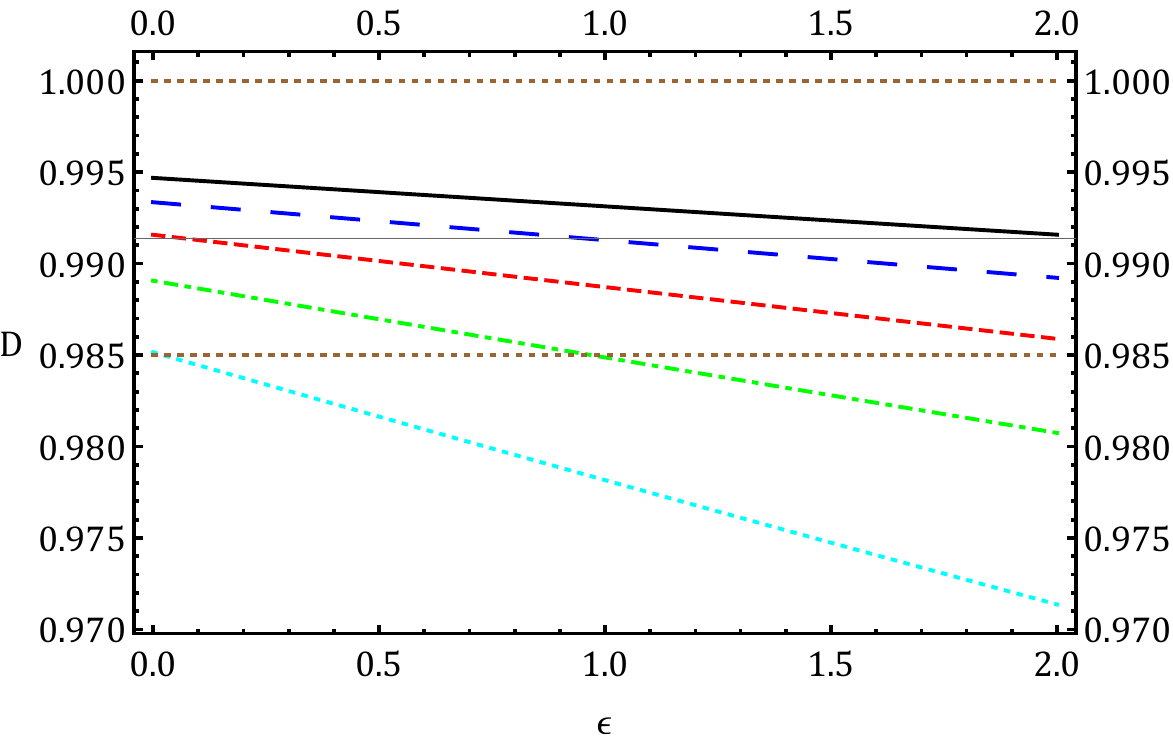}
			\caption{\textbf{Left:} Oblateness $D$ in terms of $a_*$ for different positive values of deformation parameter $\epsilon:\{0, 0.5,1,1.5, 2\}$ from the black-solid curve to cyan-dotted one, respectively. \textbf{Right:} Oblateness $D$ in terms of $\epsilon$ for several values of $a_*:\{0.8, 0.85,0.9,0.95,1\}$ from the black-solid curve to cyan-dotted one, respectively.  Brown-dotted lines denote lower and upper bounds expected from the oblateness which makes an allowed region for its change from the view of the observer with sight-angle  $\theta_0$=$17^{\circ}$.}
			\label{Oblatp}
		\end{center}	
	\end{figure}

	\begin{figure}
		\begin{center}	
			\includegraphics[width=0.6\linewidth]{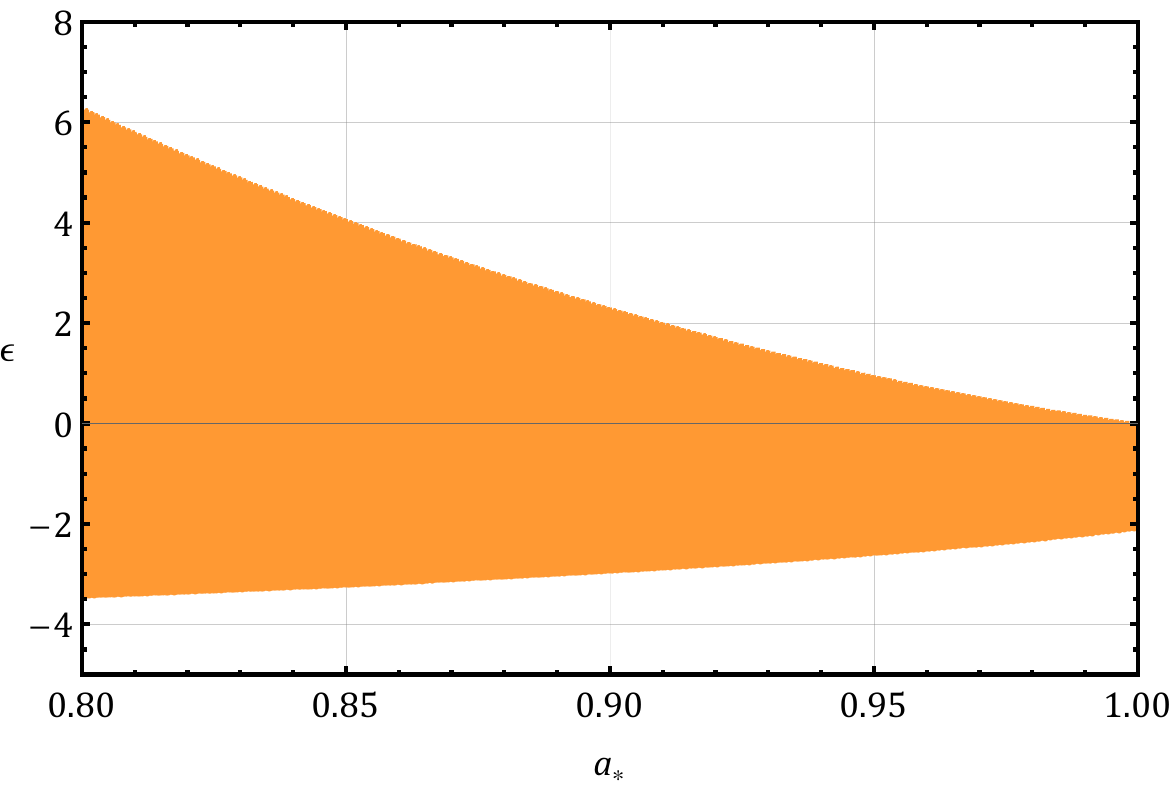}
			\caption{The permissible parameter region (colored) $a_*-\epsilon$, in which the oblateness of Kerr BH $(\epsilon<0)$ and Kerr naked singularity $(\epsilon>0)$ solutions related to the spacetime metric  (\ref{metric}), covers the expected range of $ D$ for the observer with the observation angle  $\theta_0$=$17^{\circ}$.}
			\label{psc}
		\end{center}	
	\end{figure}
	%%%%%%%%%%%%%%%%%%%%%%%%%%%%%%%%%%%%%%%%%%%%%%%%%%%%%%%%%%%%%%%%%%%
	\section{Hair parameter $\epsilon$ in the light of the EHT shadow of M87*}\label{EHT}
	%%%%%%%%%%%%%%%%%%%%%%%%%%%%%%%%%%%%%%%%%%%%%%%%%%%%%%%%%%%
	To shine a light on the additional deformation parameter $\epsilon$, we here wish to compare the shadow derived for hairy Kerr solutions at hand with the shadow of M87* SMBH \cite{Akiyama:2019cqa,Akiyama:2019eap}. This comparison performs through three observables oblateness, deviation from circularity (DC), and shadow diameter.  In this way, using the window related to these three observables, we can provide some novel upper bounds for the hair parameter $\epsilon$. It lets us evaluate the credit of NHT in the framework of the deformed Kerr spacetime metric (\ref{metric}).
		
	\subsection{Oblateness}
	The simplest observable indicating the deformation of the shadow is the oblateness \cite{Hioki:2009na,Tsupko:2017rdo,Kumar:2018ple,Neves:2019lio}. It is the ratio of horizontal and vertical diameters of the shadow, i.e. $D=\frac{\Delta\alpha}{\Delta\beta}$. The main components for computing the oblateness are the left and the right horizontal borders of shadow, $\alpha_l$, and $\alpha_r$, along with the vertical border $\beta_m$, as schematically displayed in the left panel of Fig. \ref{Shematic}. So, oblateness in principle is defined as $D=\frac{\alpha_r-\alpha_l}{2\beta_m}$. Theoretically, for the equatorial observer $(\theta_0=90^{\circ})$ the oblateness can be change from $\frac{\sqrt{3}}{2}$ to $1$ i.e. $\frac{\sqrt{3}}{2}\leq D\leq1$ in which lower and upper bounds respectively come from extreme Kerr and Schwarzschild solution \cite{Tsupko:2017rdo}. Although the upper bound is independent of the observation angle (due to spherical symmetry), the lower bound gets closer or farther from the unity, depending on the value of $\theta_0$. Actually, for the observation angles $0^{\circ}<\theta_0<90^{\circ}$, the lower bound gets closer to the unit, and the range above becomes narrower.
	In the framework at hand, there are three parameters at play: observation angle $\theta_0$, dimensionless spin parameter $a_*$, and hair parameter $\epsilon$. By fixing
	$\theta_0=17^{\circ}$ according to the amount recorded by EHT \cite{Akiyama:2019cqa,Akiyama:2019eap} as well as other observations and simulations related to measurement of the direction of outflow jets e.g. \cite{Mertens:2016rhi,Sobyanin:2018jtb,Tamburini:2019vrf}, thereby, we reveal some novel constraints on $\epsilon$ in interplay with $a_*$, see Figs. \ref{Oblatn} and \ref{Oblatp}. As can be seen from these figures, the negative and positive values of $\epsilon$ result in an increase and decrease of the oblateness relative to the standard counterpart ($\epsilon=0$), respectively. These changes in high spins are more evident. Depending on the values of spin parameter $a_*$, there are different upper bounds for the negative/positive hair parameter $\epsilon$, which can satisfy the oblateness range expected in the observation angle $\theta_0=17^{\circ}$. Here and also in what follows, due to discussions presented already about high rotation regimes, we restrict ourselves to the range $a_*=0.9\pm0.1$. In Fig. \ref{psc}, we also scanned the parameter region $a_*-\epsilon$, to find a parameter space in which the oblateness related to the deformed Kerr solutions with $\epsilon<0$ and $\epsilon>0$, be located into the theoretical allowed range. Overall,  Figs. \ref{Oblatn}, \ref{Oblatp} and \ref{psc}, contain this message that the allowed range of deformation parameter $\epsilon$ in both hairy Kerr solutions at hand (particularly $\epsilon>0$), becomes narrower, as the rotation approaches the extreme case. Actually,  by increasing the rotation parameter in the naked Kerr singularity, the upper bounds are approaching values zero.  Namely,  in the extreme limit $a_*=1$,  the hairy Kerr naked singularity is ruled out. By setting the highest value reported in the EHT paper for the dimensionless spin parameter $a_*=\mid 0.94\mid$ \cite{Akiyama:2019fyp}, one still can see an allowed narrow region of parameter space for both hairy Kerr solutions. It means a clear violation of NHT according to the present measurements of rotation parameter. Both hairy Kerr solutions become more likely if the upper bound mentioned above for the spin moves to lower values.
	However, if it moves towards the extreme case $a_*=\mid 1\mid$, the probability of ruling out hairy Kerr naked singularity becomes severe, while the hairy Kerr BH yet is at play, indicating the explicit violation of NHT.
	So, distinguish between these two hairy solutions from each other depends on more exact measurements of the spin in the future.
	
	\begin{figure}	
		\begin{center}	
			\includegraphics[width=0.5\linewidth]{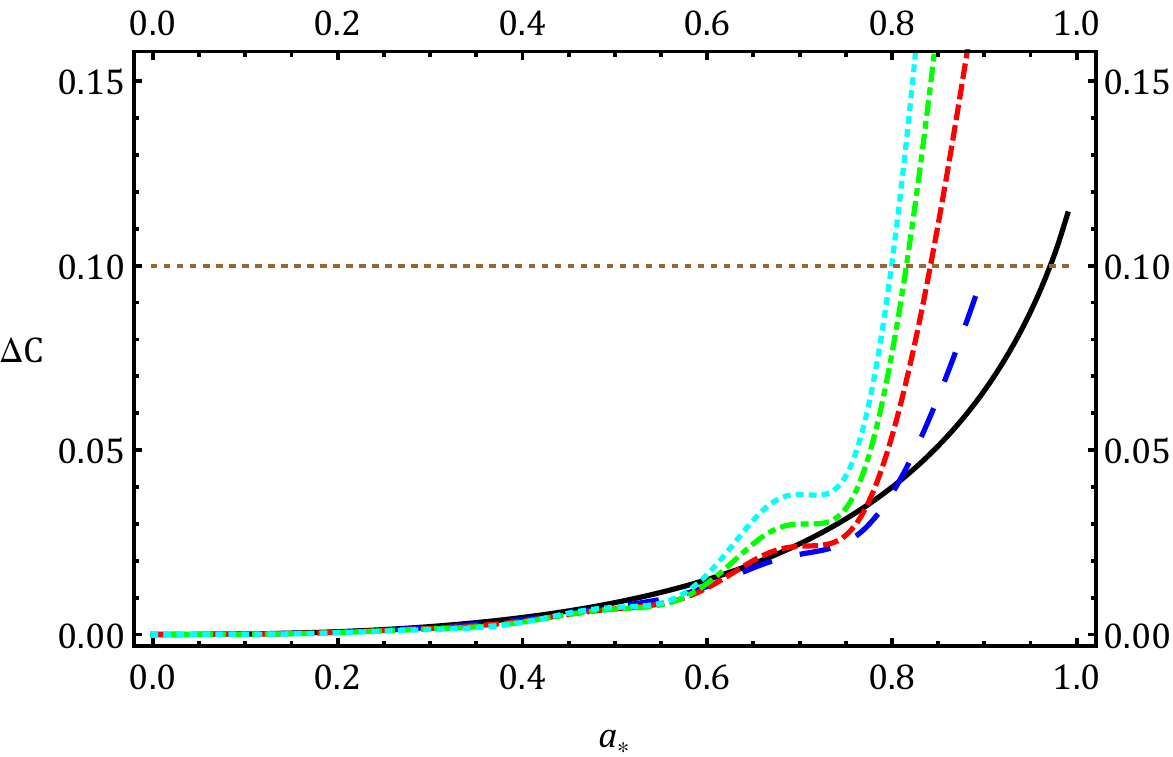}~~
			\includegraphics[width=0.5\linewidth]{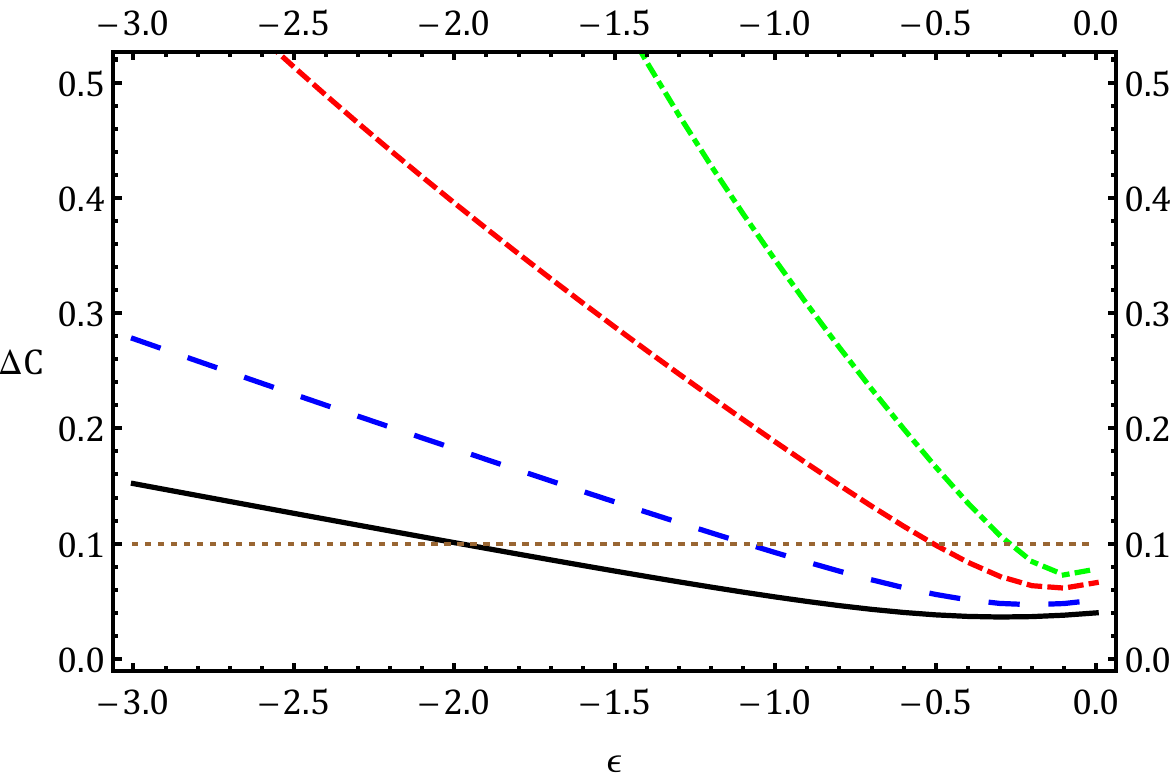}\\
			\caption{\textbf{Left:} Deviation from circularity $\bigtriangleup C$ in terms of $a_*=a/M$ for different negative values of deformation parameter $\epsilon:\{0,-0.5,-1,-1.5,-2\}$ from the black-solid curve to cyan-dotted one. \textbf{Right:}  Deviation from circularity $\bigtriangleup C$ in terms of $\epsilon$ for several values of $a_*:\{0.8, 0.85,0.9,0.95\}$ from the black-solid curve to green-dashed-dotted one, respectively.	 Brown-dotted lines denote upper bound imposed by EHT, $\bigtriangleup C\lesssim0.1$.}
			\label{DC1}
		\end{center}	
	\end{figure}
	
	\begin{figure}
		\begin{center}		
			\includegraphics[width=0.5\linewidth]{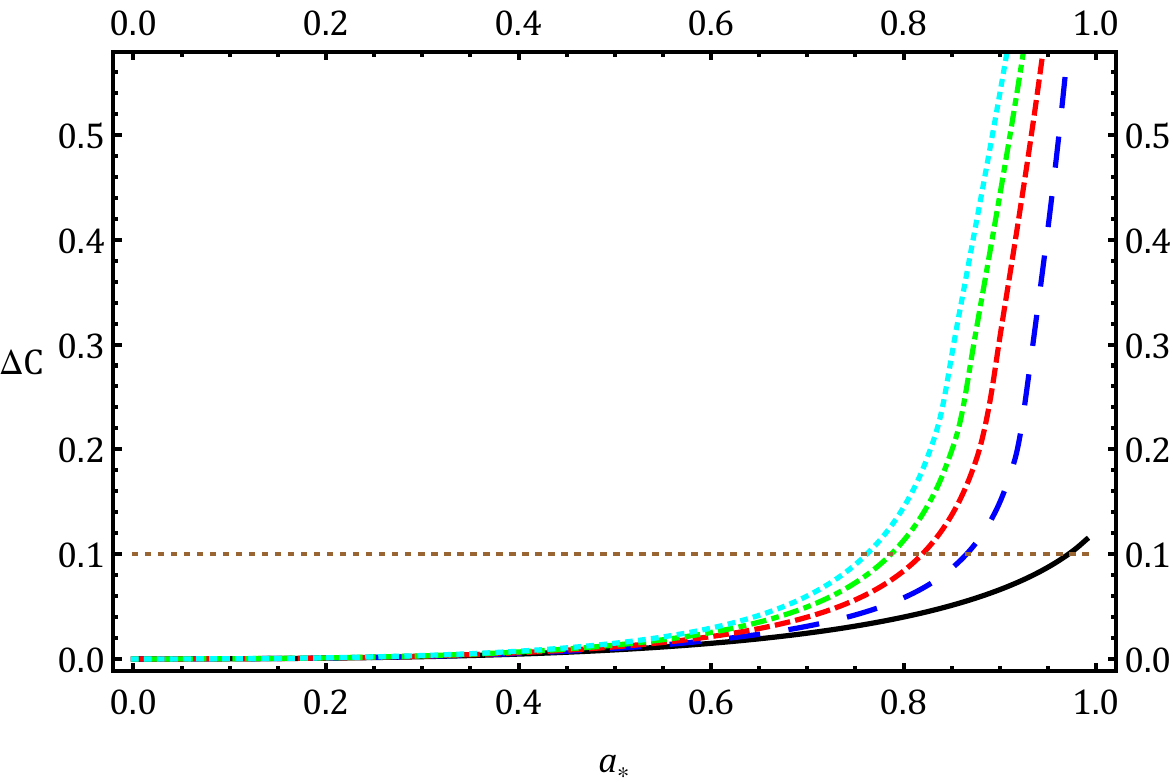}~~
			\includegraphics[width=0.5\linewidth]{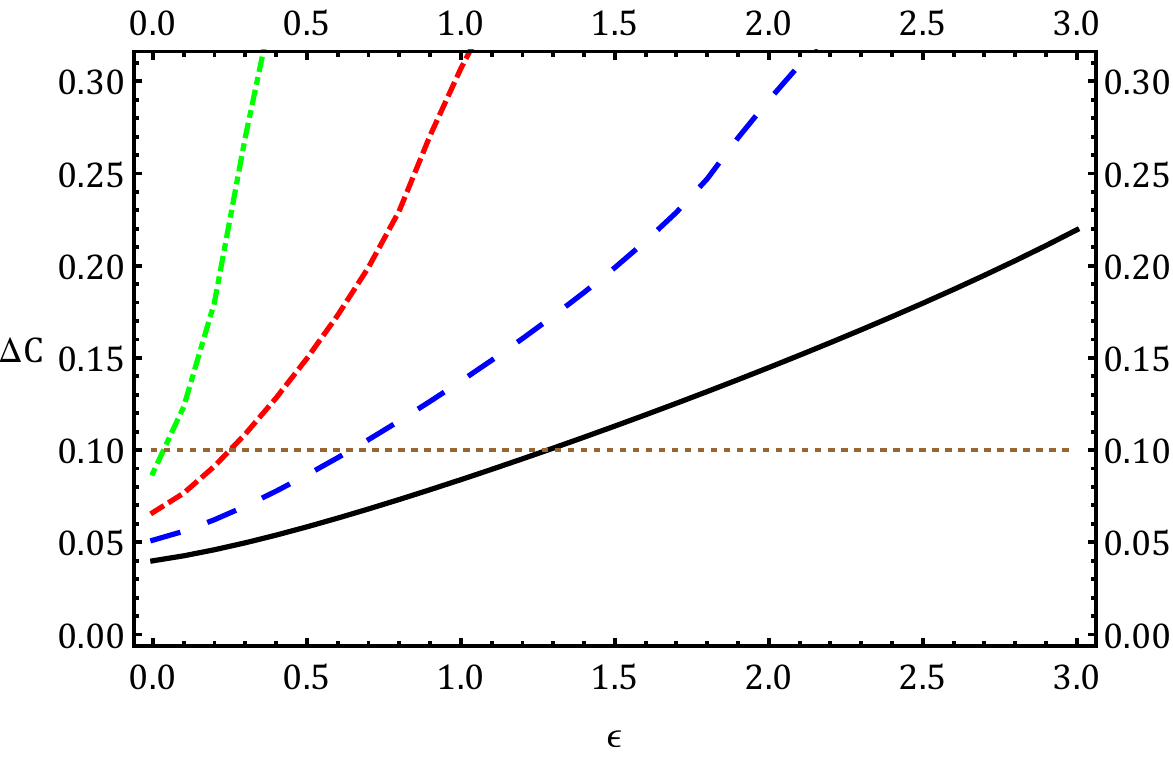}\\
			\caption{\textbf{Left:} Deviation from circularity $\bigtriangleup C$ in terms of $a_*=a/M$ for different positive values of deformation parameter $\epsilon:\{0,0.5,1,1.5,2\}$ from the black-solid curve to cyan-dotted one. \textbf{Right:}  Deviation from circularity $\bigtriangleup C$ in terms of $\epsilon$ for several values of $a_*:\{0.8, 0.85,0.9,0.95\}$ from the black-solid curve to green-dashed-dotted one, respectively.	 Brown-dotted lines denote upper bound imposed by EHT, $\bigtriangleup C\lesssim0.1$.}
			\label{DC2}
		\end{center}	
	\end{figure}
	
	\begin{figure}
		\begin{center}	
			\includegraphics[width=0.6\linewidth]{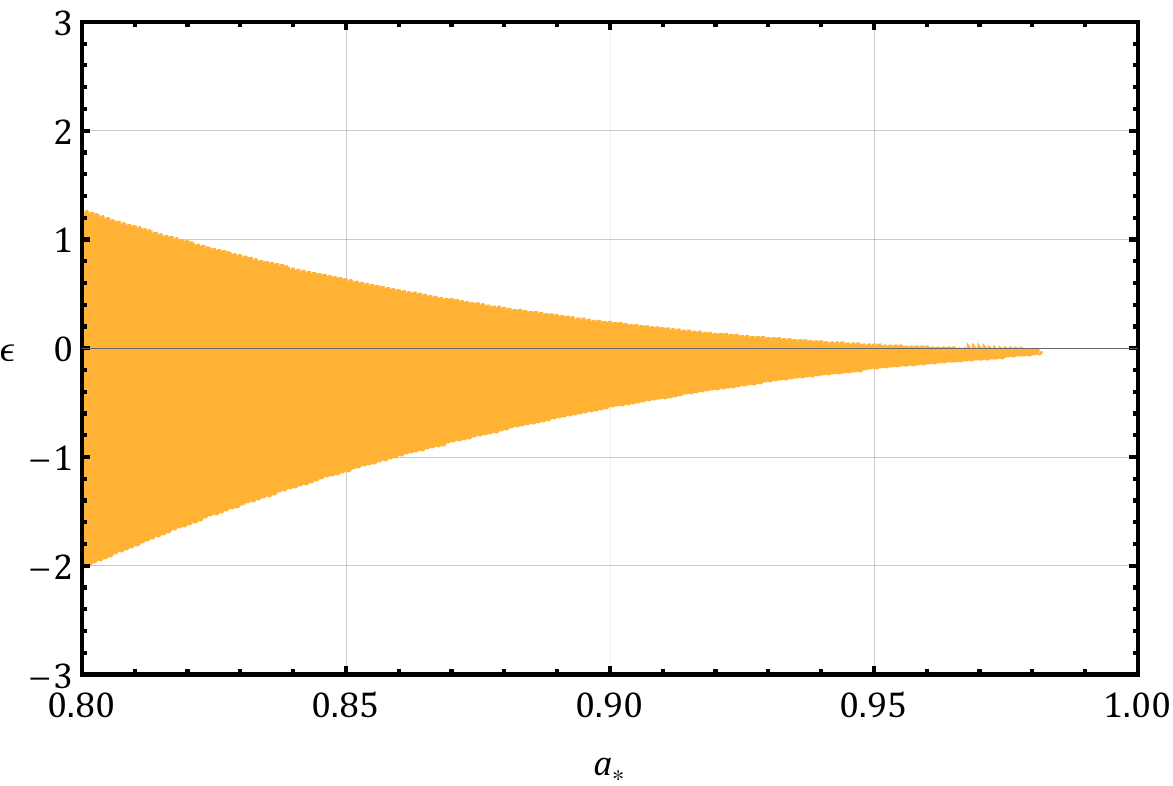}~~
			\caption{The permissible parameter region (colored) $a_*-\epsilon$, in which the DC of Kerr BH $(\epsilon<0)$ and Kerr naked singularity $(\epsilon>0)$ solutions related to the spacetime metric  (\ref{metric}), covers the allowed range of EHT, $ 0<\Delta C\lesssim 0.1$.}
			\label{pscc}
		\end{center}
	\end{figure}
	\subsection{Deviation from circularity}
	
	Based on the EHT collaboration \cite{Akiyama:2019cqa}, a criterion to address the deformation of shadow is the deviation from circularity (DC), which normally is conceived of as the deviation from the root-mean-square (rms) of the shadow radius $R_s$ and is defined as follows \cite{Bambi:2019tjh}
	\begin{eqnarray}\label{Rs}
		R_s=\sqrt{(\alpha-\alpha_c)^2+(\beta-\beta_c)^2}~,
	\end{eqnarray} where $(\alpha_c,\beta_c)$ is not the center of Cartesian coordinates, rather is the geometric center of shadow, as is evident in Fig. \ref{Shematic} (right side). Unlike oblateness, in which we deal with the left, right horizontal borders and also the vertical border of the shadow, here the main component to determine DC is the shadow radius $R_s$ whose size depending on the location of the center of the shadow $(\alpha_c,\beta_c)$. To help have a clear intuitive understanding, on the right side of Fig. \ref{Shematic}, displayed schematically the size of shadow radius $R_s$ in terms of different locations of $(\alpha_c,\beta_c)$.
The average radius (rms) is given by
	\begin{equation}\label{rms}
		\bar{R}_s=\sqrt{\frac{1}{(r_{ph^{+}}-r_{ph{-}})}\int_{r_{ph{-}}}^{r_{ph{+}}}R_s^2~ dr}.
	\end{equation}
	As already defined, here $r_{ph\mp}$ represents the radius of unstable photon sphere orbit so that the negative and positive signs are respectively a prograde and retrograde orbit parallel and anti-parallel to the direction of BH rotation.	Based on \cite{Bambi:2019tjh}, the DC actually is conceived of as the rms distance from the average radius $\bar{R}_s$, that is to say,
	\begin{equation}
		\Delta C = \sqrt{\frac{1}{(\alpha_r-\alpha_l)}\int_{r_{ph^{-}}}^{r_{ph^{+}}}\left (R_s- \bar{R}_s \right)^2 dr}.
		\label{DeltaC}
	\end{equation}
	Figs. \ref{DC1} and \ref{DC2} show that $\Delta C$ increases with the BH rotation as well as both negative and positive values of hair parameter $\epsilon$. In the EHT, assuming that the geometry of M87* supermassive BH is described by the Kerr spacetime metric, then the value of DC reported is bounded by $\Delta C \lesssim0.1$ \cite{Akiyama:2019cqa}. It can be seen from black curves in the left panels of Figs. \ref{DC1} and \ref{DC2} that from the view of the observer located in the observation angle $\theta_0=17^{\circ}$ (as imposed by EHT), the constraint $\Delta C \lesssim 0.1$ is satisfied for the standard Kerr BH ($\epsilon=0$) with spin parameters $0<a_*<1$, while in extreme value $a_*=1$, it is refuted. This statement is supported by analysis performed in \cite{Bambi:2019tjh}.
    A common feature of both figures is that by going to higher values of the BH rotation parameter $a_*$, the deformation parameter $\epsilon$ has lower chances of play a role. Actually, in the extreme case of rotation $a_*=1$, for none of the negative and positive values of $\epsilon$, there are not possible to meet the upper bound $\Delta C \lesssim0.1$. It means ruling out the hairy Kerr spacetime at hand.   In this regard, by considering both the Kerr-like objects and utilizing the M87* parameters in the DC given by Eq. (\ref{DeltaC}), one can have a more transparent understanding of constraints on the deformation parameter $\epsilon$. It is done through the full scan of parameter space in the region $a_*-\epsilon$ in interplay with the spin parameter $a_*$, see Fig. \ref{pscc}.
     This figure lets us probe the deformation parameter in detail. As we can see by fixing the highest value reported by EHT, $a_*=\mid0.94\mid$, the chance of stay at the play of the hair parameter with $\epsilon>0$  is insignificant in comparison with the case $\epsilon<0$.
	Namely, by having DC from EHT observation, one cannot conclusively rule out the non of both Kerr-like solutions expected from the deformed Kerr spacetime metric  (\ref{metric}), signaling a possible violation for NHT. However, as can be seen, the closer measurements are to the extreme limit of the rotation parameter, the more likely it is that the Kerr naked solution will rule out.
	
	\begin{figure}
		\begin{center}		
			\includegraphics[width=0.5\linewidth]{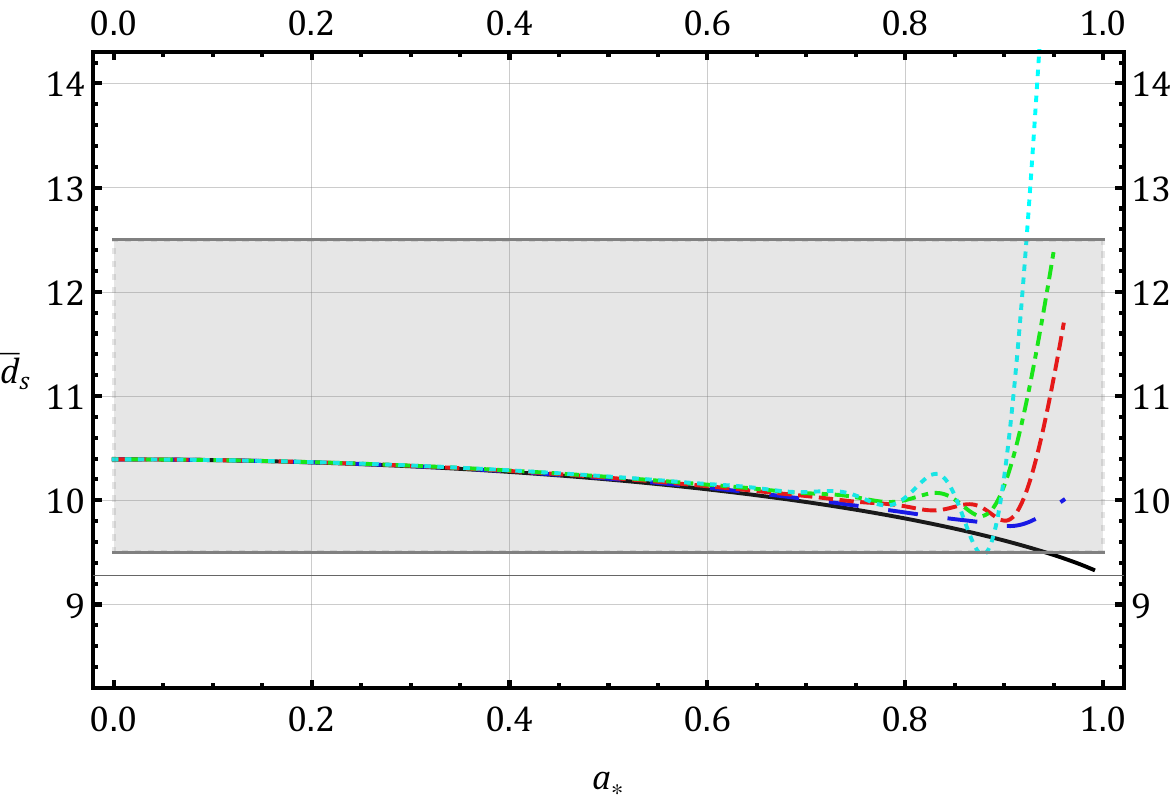}~~
			\includegraphics[width=0.5\linewidth]{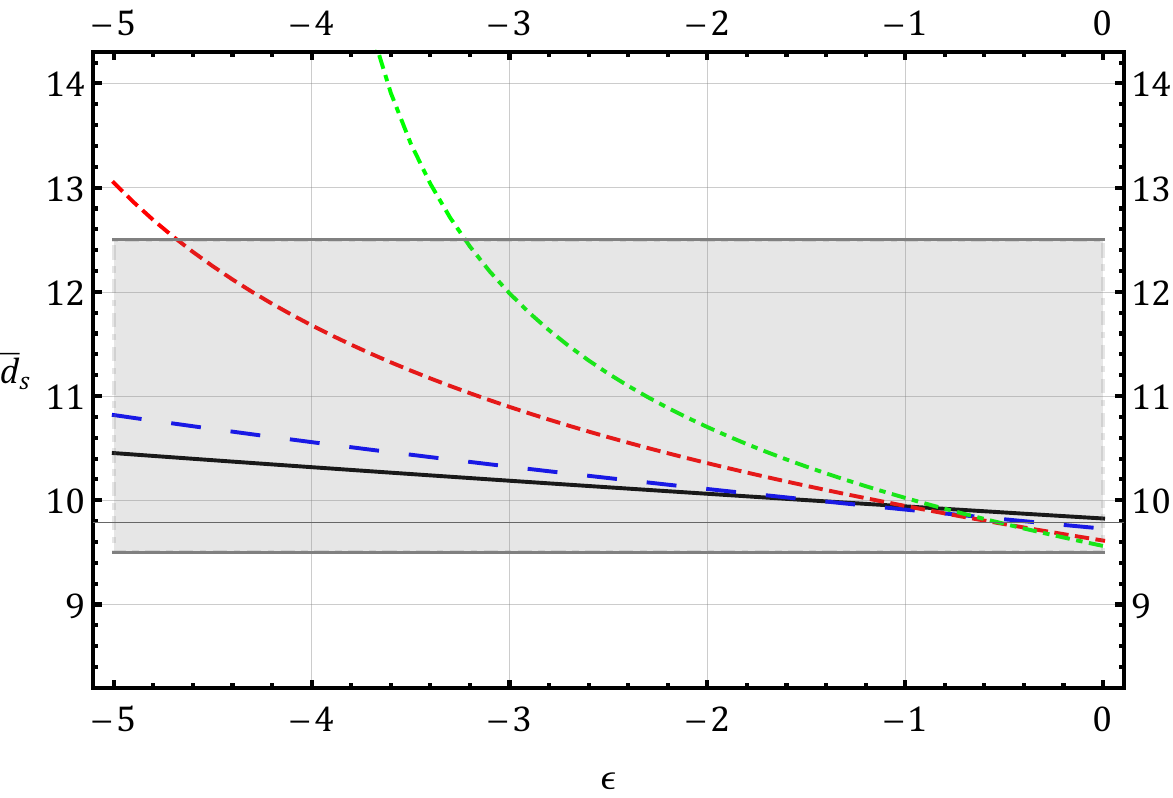}\\
			\caption{\textbf{Left:} Shadow diameter $\bar{d}_s$ in terms of $a_*$ for different positive values of deformation parameter $\epsilon:\{0, -0.5,-1,-1.5,-2\}$ from the black-solid curve to cyan-dotted one. \textbf{Right:} Shadow diameter $\bar{d}_s$ in terms of $\epsilon$ for several values of $a_*:\{0.8, 0.85,0.9,0.95\}$ from the black-solid curve to green-dashed-dotted one, respectively.}
			\label{ShR}
		\end{center}	
	\end{figure}
	
	\begin{figure}
		\begin{center}		
			\includegraphics[width=0.5\linewidth]{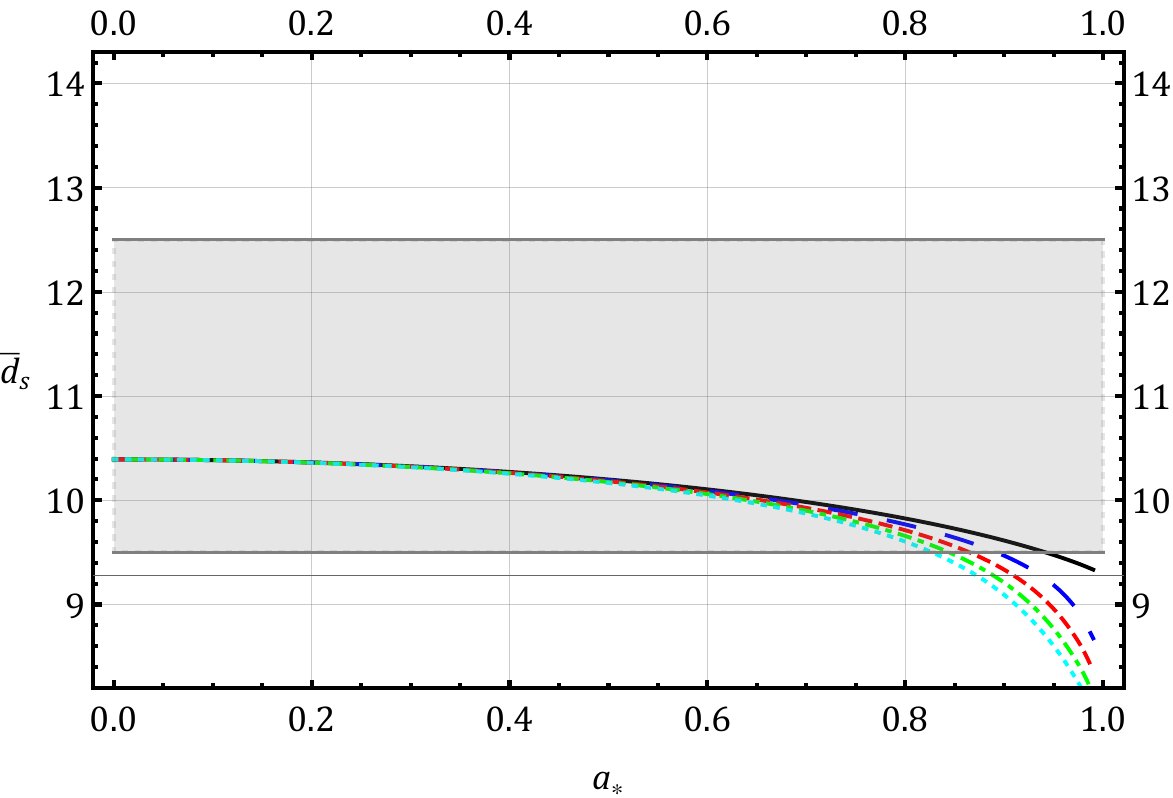}~~
			\includegraphics[width=0.5\linewidth]{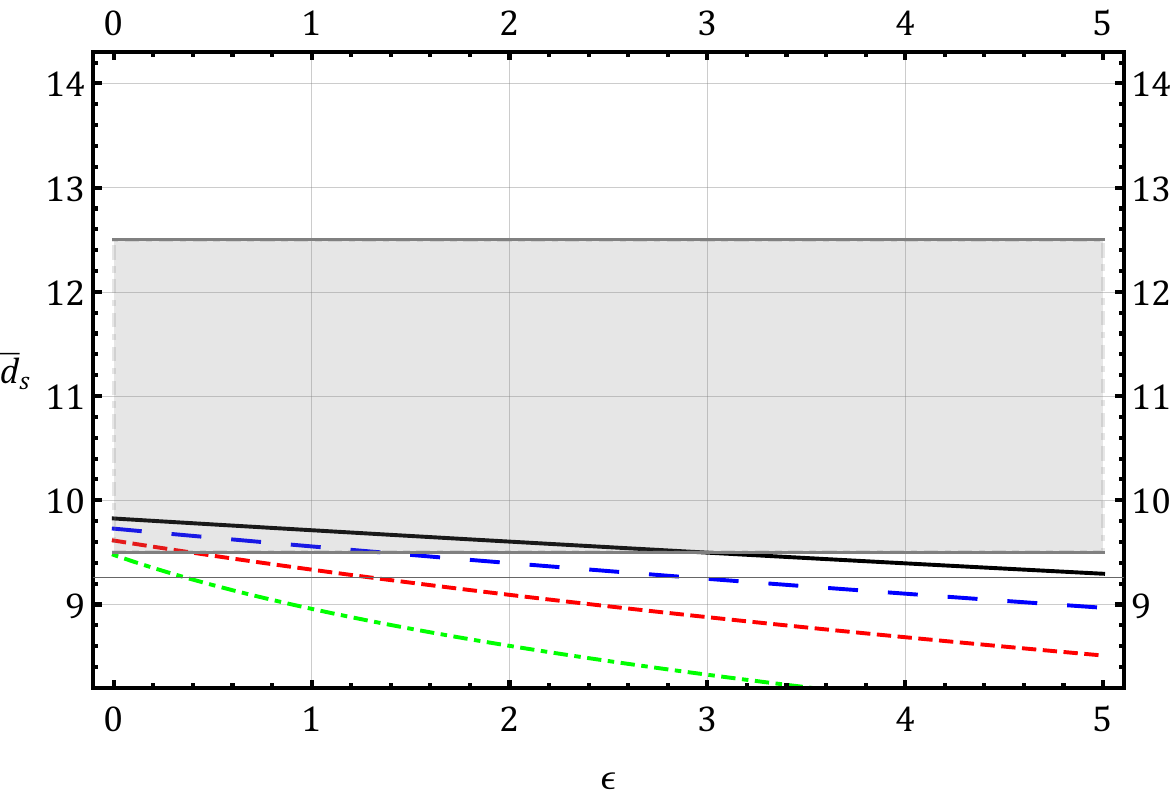}\\
			\caption{\textbf{Left:} Shadow diameter $\bar{d}_s$ in terms of $a_*$ for different positive values of deformation parameter $\epsilon:\{0, 0.5,1,1.5,2\}$ from the black-solid curve to cyan-dotted one. \textbf{Right:} Shadow diameter $\bar{d}_s$ in terms of $\epsilon$ for several values of $a_*:\{0.8, 0.85,0.9,0.95\}$ from the black-solid curve to green-dashed-dotted one, respectively.}
			\label{ShRR}
		\end{center}	
	\end{figure}
	
	\begin{figure}
		\begin{center}	
			\includegraphics[width=0.6\linewidth]{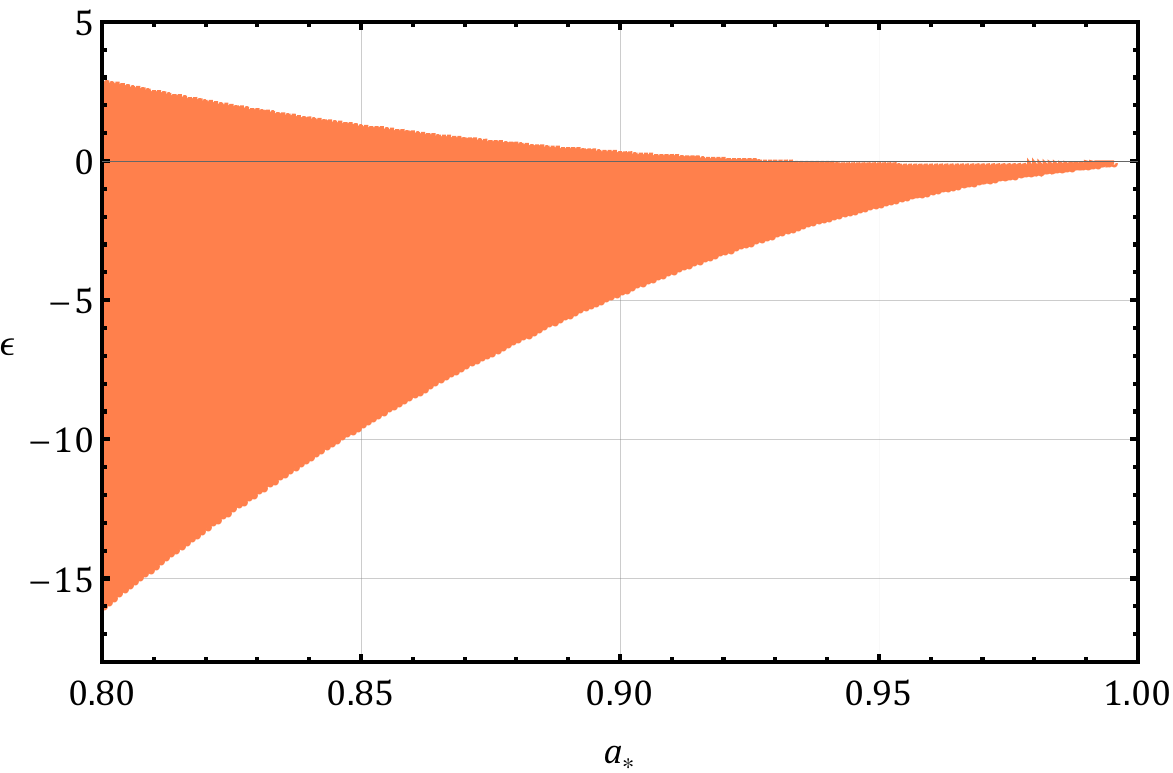}~~
			\caption{The permissible parameter region (colored) $a_*-\epsilon$, in which the diameter in units of mass of Kerr BH $(\epsilon<0)$ and Kerr naked singularity $(\epsilon>0)$ solutions related to the spacetime metric  (\ref{metric}), covers the observational values estimated for the M87* within $1\sigma$ uncertainty, $9.5\lesssim d_{M87*}\lesssim12.5$.}
			\label{ps}
		\end{center}	
	\end{figure}
	
	\subsection{Shadow diameter}
	
	As we saw earlier in Figs.~\ref{Sh1} and \ref{Sh2}, the shadow shape depends rather strongly on the value of the hair parameter $\epsilon$. Therefore, here we want to utilize the shadow diameter as another observable involved in the observation of M87* by EHT to probe hairy Kerr solutions expected from spacetime metric (\ref{metric}). Following the information released in \cite{Akiyama:2019cqa,Akiyama:2019eap}, one finds values $\delta = (42 \pm 3)\,\mu{\rm as}$, $D = 16.8^{+0.8}_{-0.7}\,{\rm Mpc}$
	and  $M = (6.5 \pm 0.9) \times 10^9\,M_{\odot}$ corresponding to the angular size of the shadow of M87*, the distance
	to M87*  and the mass of M87*, respectively.
	By merging  these data, the diameter of shadow in units of mass $d_{M87*}$, reads as \cite{Bambi:2019tjh}
	\begin{eqnarray}
		d_{M87*} \equiv \frac{D\delta}{M} \approx 11.0 \pm 1.5\,.
		\label{size}
	\end{eqnarray}
	As a notable point, the detected diameter of M87* shadow, as given in Eq.~(\ref{size}), is compatible with that of the
	Schwarzschild BH. This can be verified via setting the $\bar{d}_s=2 \bar{R}_s$ within Eq.~(\ref{size}), see black-solid curves with $\epsilon=0$ in $a_*=0$ in the left panels of Figs.~\ref{ShR} and~\ref{ShRR} which analyzed within $1\sigma$-uncertainty  $9.5\lesssim d_{M87*}\lesssim12.5$. Note that here $\bar{R}_s$ is the average of shadow radius and comes from (\ref{rms}). These two figures clearly show that by turning on the deformation parameters $\epsilon<0$ and $\epsilon>0$, the shadow diameter $\bar{d}_s$  respectively grows and compacts by increasing the dimensionless spin parameter $a_*$. Actually, by restricting	the growth or reduce of the shadow within the range $9.5\lesssim d_{M87*}\lesssim 12.5$, one can extract some novel bounds for the hair parameter $\epsilon$. In this regard, we do a scan of the parameter region $a_*-\epsilon$, as can be seen in Fig.~\ref{ps}. Interestingly, we can see that by nearing the extreme limit $a_*=1$, the Kerr naked singularity solution is refuted, while its BH counterpart yet is at play, indicating the violation of NHT.
	%%%%%%%%%%%%%%%%%%%%%%%%%%%%%%%%%%%%%%%%%%%%%%%%%%%%%%%%%%%%%%%%%%%
	\section{Discussion and Conclusion}\label{final}
	
The first super-captured image of the M87* BH shadow by the EHT team made us no longer think of these strange objects as fantasies. Actually, now more than ever this compact object imagine as a perceptible reality that lets us perform the strong-field tests of the GTR and fundamental physics. Data analysis released by EHT collaboration indicates that the standard Kerr metric can well-describe the M87* BH shadow. However, this does not mean the exclusion of other classes of BH solutions that fall down within the window opened by
EHT measurements of M87*, and therefore there can still be a chance for them to stay in the competition.
	
	In this way, as a well-behaved framework to describe of M87* BH shadow, we have adopted the deformed Kerr metric (\ref{metric}) proposed by Johannsen \& Psaltis that is characterized by a dimensionless deformed (or hair) parameter $\epsilon$ in addition to standard parameters of mass and spin.  So, through the EHT measurements of M87*, one can perform a strong-gravity test of one of the fundamental issues in physics i.e. the no-hair theorem (NHT). The interesting  property of the deformed Kerr metric (\ref{metric}) is that it has twofold implications: Kerr-like naked singularity and Kerr-like BH solutions, depending on setting positive and negative values for $\epsilon$, respectively.  As a valuable feature in the Kerr-like BH solution, up to the maximum values of the dimensionless spin parameter $a_*=1$, everywhere outside the region of the event horizon is regular and free of any pathological behaviors. However, because the naked singularity is also able to make a shadow, hence for the test of NHT in the strong-gravity limit, both Kerr solutions have been confronted to the EHT measurements of M87* BH shadow.
Actually, in the context of the metric at hand, we have two alternative solutions to the standard Kerr, and both can be tested through EHT. 
An important point to note is that our analysis is restricted to high rotation regimes, in particular in the range $a_*=0.9\mp 0.1$. Apart from the theoretical popularity of fast-rotating BHs for the reasons already discussed, the mentioned range comes from the simulations done for the twist of the light emitted from the Einstein ring surrounding the M87* BH shadow in EHT observation angle $\theta_0=17^{\circ}$. The data analysis of the EHT team has also indicated that BHs have high rotation.  Below is a summarized report of performed analysis as well as results obtained in this paper.
	
	\begin{itemize}
		\item First of all, we have analyzed how the presence of the hair parameter $\epsilon$, affects the BH shadow.
		The curves revealed in Figs.~\ref{Sh1}, and \ref{Sh2} clearly show that by increasing negative (module) and positive values of $\epsilon$, thereby, the shape of BH shadow become more oblate and prolate relative to standard Kerr, respectively. 
		This means that both negative and positive values of deformation parameter $\epsilon$ can affect the geometry shape of BH shadow. Note that the amount of this deformation in addition to $\epsilon$ depends on the dimensionless rotation parameter $a_*$ and the observer angle of view $\theta_0$, too.
		\item In the first step to applying constraints on $\epsilon$ associated with hairy Kerr spacetime, we have taken the simplest observable related to the distortion of shadow, oblateness $D$. The effects of oblateness induced by the deformation parameter $\epsilon$ are such that in the presence of negative and positive values, $D$ increases and decreases relative to the standard counterpart ($\epsilon=0$), respectively. By scanning the allowed window of $D$ in parameter space of the region $a_*-\epsilon$, we have extracted the range $-4<\epsilon\leq6$ within the demanded range of the rotation parameter. We found that by increasing the rotation toward the extreme case $a_*=1$, the allowed range of the deformation parameter $\epsilon$ in both hairy Kerr solutions at hand (particularly $\epsilon>0$), become narrower so that for the case of naked Kerr singularity the upper bounds are approaching zero. Namely, in the extreme limit of the rotation parameter, there is the possibility of ruling out the hairy Kerr naked singularity.  Even if one sets the highest value reported by EHT collaboration for the dimensionless spin parameter $a_*=\mid 0.94\mid$, there is still a narrow range of parameter space for the hairy Kerr solutions that match the allowed window of oblateness $D$. This means a clear violation of NHT according to the present measurement of the spin parameter. Note that the violation exists in any case because the hairy Kerr BH solution remains in the play under any constraint on the dimensionless spin parameter $a_*$. The above results can be seen in Figs. \ref{Oblatn}, \ref{Oblatp} and \ref{psc}.
		\item As the next step, we picked up another shadow observable, the deviation from circularity (DC) $\bigtriangleup C$, which, according to the EHT limit, varies in  the range $0<\bigtriangleup C\lesssim 0.1$. We first have shown that both negative and positive values of hair parameter $\epsilon$ in interplay with the BH rotation parameter $a_*$, affects $\Delta C$. More exactly, by increasing rotation, the chance of satisfying the above-mentioned allowed range for both hairy Kerr solutions comes down. Our full scan of parameter space in the region $a_*-\epsilon$ has demonstrated an allowed range $-2\leq \epsilon<1.5$ within the desired range of the dimensionless spin parameter. The analysis of the parameter space has revealed that in high rotation regimes, in particular by taking the highest value reported by EHT, $a_*=\mid 0.94\mid$, there is more chance of the hair parameter surviving for the case of $\epsilon<0$ relative to $\epsilon>0$. In other words, by having DC arising from EHT observation, one cannot conclusively refute any of the possible solutions expected from the deformed Kerr spacetime metric  (\ref{metric}), signaling the possible violation for NHT. These results are traceable via Figs. \ref{DC1}, \ref{DC2} and \ref{pscc}.
		\item Ultimately, we have utilized the M87* SMBH shadow diameter $d_{M87*}$, to find upper bounds on the deformation parameter $\epsilon$. We found that the average shadow diameter $\bar{d}_s$ of hairy Kerr solutions at hand grows and reduces respectively in terms of the deformation parameters $\epsilon<0$ and $\epsilon>0$, as the dimensionless spin parameter $a_*$, increases.  Through the full scan of parameter space in the region $a_*-\epsilon$, we have extracted the allowed range $-16<\epsilon<3$ within the demanded values of $a_*$.  This result rejects the possibility of the existence of a hairy Kerr naked singularity solution ($\epsilon>0$) with rotation parameter close to extreme value $a_*=1$. However, it has no conflict with the hairy Kerr BH solution ($\epsilon<0$).
		By taking $a_*=\mid 0.94\mid$ as the highest value admitted by EHT for the rotation parameter, it has shown that the hairy Kerr naked singularity solution has no chance of survival.
		The above results also can be seen in Figs. \ref{ShR}, \ref{ShRR} and \ref{ps}.
	\end{itemize}
	Overall, the most solid constraint for the hair parameter $\epsilon$ among the three cases mentioned above belongs to DC. However, from the perspective of the distinction between these two deformed Kerr-like solutions, the most interesting result comes from the last case in which the EHT data conclusively rules out the existence of naked Kerr singularity.
	In all results obtained for the three underlying shadow observables, there are two common points. First, the explicit violation of NHT, in the light of the current observation of EHT. Second, dependency of all the above results on the value of the dimensionless rotation parameter $a_*$. In other words, more exact measurements of the rotation parameter in the future will shine a light on these results.

%%%%%%%%%%%%%%%%%%%%%%%%%%%%%%%%%%%%%%%%%%%%%%%%%%%%%%%%%%%%%%%%%%%
\vspace{1cm}
{\bf Acknowledgments:}
MKh would like to thanks Alireza Talebian for the helpful comments. GL thanks INFN and MIUR for support. DFM thanks the Research Council of Norway for their support. Computations were performed on resources provided by UNINETT Sigma2 -- the National Infrastructure for High Performance Computing and Data Storage in Norway.
	
	\providecommand{\href}[2]{#2}\begingroup\raggedright
	
\end{document}